\documentclass[11pt]{article}
\pdfoutput=1

\usepackage{amstext,amsmath,amssymb,amsfonts,amsthm}
\usepackage{bbm}
\usepackage{hyperref}
\usepackage{authblk}
\usepackage{caption}
\usepackage[latin1]{inputenc}
\usepackage{subfigure}
\usepackage{color}
\usepackage{graphicx}
\usepackage{braket}
\usepackage[nospace,nosort]{cite} % for compressing lists of refs (e.g. [1-3] instead of [1,2,3]), and more

\usepackage[lmargin=80pt,rmargin=80pt,tmargin=80pt,bmargin=80pt]{geometry}

\numberwithin{equation}{section}

\newcommand{\be}{\begin{equation}}
\newcommand{\ee}{\end{equation}}
\newcommand{\nn}{\nonumber}
\newcommand{\f}{\frac}
\newcommand{\p}{\partial}

\newcommand{\Tr}{{\rm Tr}}
\newcommand{\tr}{{\rm tr}}

\newcommand{\la}{\langle}
\newcommand{\ra}{\rangle}
\newcommand{\dd}{{\rm d}^d}
\newcommand{\tj}[6]{ \begin{pmatrix}
       #1 & #2 & #3 \\
       #4 & #5 & #6 
\end{pmatrix}}

\newtheorem{proposition}{Proposition}
\newtheorem*{proposition*}{Proposition}

\newtheorem{claim}{Claim}
\newtheorem{conjecture}{Conjecture}
\newtheorem{hypothesis}{Hypothesis}
\theoremstyle{remark}

\DeclareMathOperator{\im}{\mathrm{i}}

\let\a=\alpha     \let\d=\delta
\let\z=\zeta        \let\l=\lambda
\let\m=\mu    \let\n=\nu          \let\r=\rho 
\let\s=\sigma \let\t=\tau    \let\ph=\phi 
    
\let\G=\Gamma \let\D=\Delta    \let\X=F
         
  \let\eps=\epsilon

\newcommand{\mba}{\mathbf{a}}
\newcommand{\mbb}{\mathbf{b}}
\newcommand{\mbc}{\mathbf{c}}
\newcommand{\mbd}{\mathbf{d}}

\newcommand{\htilde}{\tilde{h}}

\newcommand{\tG}{\tilde{G}}

\newcommand{\wtD}{\widetilde{\Delta}}
\newcommand{\wtX}{\widetilde{X}}

\newcommand{\cF}{\mathcal{F}}
\newcommand{\cG}{\mathcal{G}}

\newcommand{\cJ}{\mathcal{J}}

\newcommand{\cN}{\mathcal{N}}
\newcommand{\cO}{\mathcal{O}}
\newcommand{\cP}{\mathcal{P}}

\newcommand{\cS}{\mathcal{S}}

\newcommand{\cZ}{\mathcal{Z}}

\begin{document}

\title{\bf Instability of complex CFTs\\ with operators in the principal series}

\author{Dario Benedetti}
%\author[2]{...}

\affil{\normalsize\it CPHT, CNRS, Ecole Polytechnique, Institut Polytechnique de Paris, Route de Saclay, \authorcr \it 91128 Palaiseau, France \authorcr
email: {dario.benedetti@polytechnique.edu}
  \authorcr \hfill }

%\affil[2]{\normalsize\it ... }

\date{}

\maketitle

\hrule\bigskip

\begin{abstract}

We prove the instability of $d$-dimensional conformal field theories (CFTs) having in the operator-product expansion of two fundamental fields a primary operator 
 of scaling dimension $h=\f{d}{2}+\im  r$, with non-vanishing $ r\in\mathbb{R}$.
From an AdS/CFT point of view, this corresponds to a well-known tachyonic instability, associated to a violation of the Breitenlohner-Freedman bound in AdS${}_{d+1}$; we derive it here directly for generic $d$-dimensional CFTs that can be obtained as limits of multiscalar quantum field theories, by applying the harmonic analysis for the Euclidean conformal group to perturbations of the conformal solution in the two-particle irreducible (2PI) effective action.
Some explicit examples are discussed, such as melonic tensor models and the biscalar fishnet model.

\end{abstract}

\bigskip\hrule\bigskip

%\newpage
\tableofcontents
%\newpage

%%%%%%%%%%%%%%%%%%%%%%%%%%%%%
\section{Introduction and summary}
%%%%%%%%%%%%%%%%%%%%%%%%%%%%%

All the correlators of local operators in a conformal field theory (CFT) are fully determined by its CFT data, that is, the spectrum of scaling dimensions of primary operators, and the operator-product expansion (OPE) coefficients. 
Obtaining estimates of possible CFT data by tightly bounding them with sets of self-consistency equations is the main goal of the conformal bootstrap program, which has achieved impressive results \cite{Poland:2018epd}.
The simplest constraints are those provided by unitarity, which is essential in high-energy physics applications, and which demands that CFT data are real, and the scaling dimensions are above unitarity bounds. However, unitarity (or reflection positivity) is not a necessary requirement in statistical physics applications, hence it seems legitimate to consider Euclidean CFTs violating the unitarity bounds, and perhaps even the reality constraints. Moreover,  complex CFTs have also been related to walking behavior in  real and unitary theories \cite{Gorbenko:2018ncu}, showing that their presence in an extended theory space can be relevant even in high-energy physics.

One natural way in which CFTs with complex scaling dimensions can arise is as renormalization group fixed points at complex values of the couplings, for example when real fixed points merge and move to the complex plane as some parameter of theory is varied. 
Many examples of such merging of fixed points are known, see for example the list of references in \cite{Faedo:2019nxw}.
In some instances, a special case of complex scaling dimension is found, one whose real part is equal to $d/2$,  $d$ being the space(-time) dimension; examples include non-supersymmetric orbifolds of $\cN = 4$ super Yang-Mills \cite{Dymarsky:2005uh,Dymarsky:2005nc,Pomoni:2008de}, gauge theories with matter in the Veneziano limit \cite{Kaplan:2009kr,Benini:2019dfy}, and large-$N$ theories dominated by melonic diagrams \cite{Murugan:2017eto,Giombi:2017dtl,Prakash:2017hwq,Giombi:2018qgp,Kim:2019upg,Benedetti:2019eyl,Benedetti:2019rja,Klebanov:2020kck,Benedetti:2020iku}, or by fishnet diagrams \cite{Grabner:2017pgm,Kazakov:2018qbr,Gromov:2018hut,Kazakov:2018gcy,Pittelli:2019ceq,Levkovich-Maslyuk:2020rlp}.
The typical large-$N$ mechanism leading to complex dimensions of such type is the following \cite{Pomoni:2008de}: due to the large-$N$ simplifications, the beta function for the coupling of a double-trace operator $\cO^2$ turns out to be governed by a quadratic beta function, and hence the reality of the fixed points depends on its discriminant $D$; in the case $D<0$, the fixed points are complex, and the scaling dimension of $\cO^2$ is $\D_{\cO^2}=d+  2 \im \sqrt{|D|}$; lastly, the large-$N$ limit implies that the scaling dimension of $\cO$ is half that of $\cO^2$, and thus it is of the claimed form.
Here we will be concerned with such particular case of scaling dimensions, although not necessarily in the large-$N$ limit.

Operators with a scaling dimension equal to $d/2+\im r$ (with $r\in\mathbb{R}$) are commonly seen as a sign of instability,
the reason being that in  the AdS/CFT correspondence they correspond to tachyonic fields. By the standard AdS/CFT dictionary \cite{Gubser:1998bc,Witten:1998qj} the conformal dimension $\D$ of a scalar operator  in the $d$-dimensional CFT is related to the mass $m$ of a scalar field in $(d+1)$-dimensional anti-de Sitter space (AdS${}_{d+1}$) by the equation $m^2= \D (\D-d)$, or
\be \label{eq:Delta-m}
\D_\pm=\f{d}{2} \pm \sqrt{\f{d^2}{4}+m^2} \,.
\ee
In AdS${}_{d+1}$, thanks to the constant curvature of spacetime, the squared mass can be negative without generating a tachyonic instability, as long as $m^2\geq - d^2/4$. The latter is the well-known Breitenlohner-Freedman bound \cite{Breitenlohner:1982bm,Breitenlohner:1982jf}, below which a tachyonic instability develops.
The square root in\eqref{eq:Delta-m} becomes imaginary precisely  for $m^2< - d^2/4$, hence the expectation that CFTs with such complex scaling dimensions are unstable, in the sense that the conformal vacuum is not the true vacuum of the theory.\footnote{Within the large-$N$ setting discussed above, Pomoni and Rastelli in \cite{Pomoni:2008de} have shown that the theory with real coupling is in a broken phase; however, they have not directly argued for the instability of the fixed point theory itself, which is at complex coupling, instead referring for that to the AdS/CFT picture.}
However, elevating this argument to a proof does not seem feasible, in particular because the AdS/CFT correspondence is mostly based on matching calculations on both sides of the conjecture, something that requires having at least a guess for the bulk dual of a given CFT, while in some of the models displaying  a scaling dimensions equal to $d/2+\im r$ this is currently missing.
Moreover, most of such matching calculations rely importantly on the large-$N$ limit, while one might suspect that if a complex scaling dimension of this type is a synonym of instability, this might be a more general result. The reason is that $d/2+\im r$ also happens to be the conformal dimension labelling the principal series representations of the Euclidean conformal group $SO(d+1,1)$, and therefore it seems plausible that we might not need the large-$N$ limit in order to single out such scaling dimensions. And although we only know of examples involving a large-$N$ limit, this might be just because for either fundamental or practical (i.e.\ computational) reasons the limit is essential for the appearance of such type of scaling dimensions in a CFT, and not because it is needed in order to prove the instability. 

In this paper we will construct a proof of such instability directly for generic $d$-dimensional CFTs that can be obtained as limits of multiscalar quantum field theories, without invoking the large-$N$ limit, and of course under a set of assumptions that will be specified later. 
The main result of the paper is informally summarized by the following statement:

\begin{claim}\label{claim1} Consider a Euclidean quantum field theory whose Schwinger-Dyson equations admit a conformal solution. If the OPE of two fundmental scalar fields includes a contribution from one primary operator $\cO_{h_\star}$ of dimension $h_\star=\f{d}{2}+\im  r_\star$, with non-vanishing $ r_\star\in\mathbb{R}$, then the conformal solution is unstable.
\end{claim}

By fundamental fields we here mean the fields which enter the definition of the quantum field theory, through the functional integral.
The setting and precise statement of the hypotheses and conclusion will be made more sharp in Section~\ref{sec:proof}, culminating in Proposition~\ref{prop1}.
The proof is not fully rigorous in the mathematical physics sense, because we bypass all the issues of renormalization of the underlying quantum field theory, in the effort to keep the treatment as general as possible, without restricting to a specific model. We assume that the CFT can be obtained as a fixed point of the renormalization group, and that at least in some limit the conformal two-point functions of the fundamental scalar fields can be identified as solutions of the Schwinger-Dyson (SD) equations. The latter point is where in practice the large-$N$ limit (often together with an IR limit) plays a crucial role, as without it we are usually not able to solve the SD equations. However, this does not seem to be a fundamental requirement, as in principle the SD equations should admit a solution anyway.
The main claim then is that a conformal solution leading to a primary operator of scaling dimension $h_\star=\f{d}{2}+\im  r_\star$ is necessarily unstable.
By unstable here we mean that the free energy is not minimized (not even locally) by the conformal solution of the SD equations, as there exist fluctuations that lower the free energy.
 
In order to further clarify the meaning of instability in this context, it might be useful to compare the type of instability we are discussing here with a more familiar one. Consider a single-scalar field theory in $d\geq 2$ with quartic interaction and $\mathbb{Z}_2$ invariance. By tuning the renormalized mass, the model can go through a phase transition. One way to see that is by constructing the effective potential $V(\phi)$ (i.e.\ the one-particle irreducible (1PI) effective action at constant field configuration, divided by the volume), whose stationary point $V'(\phi_\star)=0$ gives the one-point function, $\phi_\star=\la \varphi \ra$, while $V(\phi_\star)$ is the free energy per unit volume in the absence of external sources.
At negative squared mass, we find that the symmetric solution $\phi_\star =0$ has become a local maximum of the potential, and that non-trivial minima have appeared, with a lower free energy. In this case, we say that the symmetric solution is unstable.
Similarly, in the case of Claim~\ref{claim1}, we want to show that the conformal solution of the SD equations is not a (local) minimum of the free energy.\footnote{A very explicit example of instability of this kind is provided by the Bardeen-Moshe-Bander phenomenon \cite{Bardeen:1983rv} (see also \cite{Amit:1984ri,Omid:2016jve,Marchais:2017jqc,Fleming:2020qqx} and references therein), which however is not associated to complex scaling dimensions.}
We will not have any claim instead on the existence and properties of other solutions, except for some small remarks in the concluding section. In particular, the question of whether such models admit a stable solution with spontaneous breaking of conformal invariance, or they are fully unstable and pathological, might be model-dependent, and it is left open.

We outline here a sketch of the proof, hopefully useful for the rushed reader, but also to highlight the gist of the lengthier proof in Section~\ref{sec:proof}, whose main ideas might be lost among the technicalities.
The sketch of the proof goes as following:
\begin{enumerate}
\item We use the two-particle irreducible (2PI) effective action formalism (see for example  \cite{Cornwall:1974vz,Berges:2004yj,Benedetti:2018goh}), in which field equations for the effective action are the Schwinger-Dyson equations for the two-point function $G(x,y)$ of the fundamental fields $\phi(x)$; this is the two-point function analogue of the usual effective potential for the one-point function.
The first main hypothesis consists in assuming that such equations admit a conformal solution, $G_\star(x,y)\sim1/|x-y|^{2\D}$.

\item The 2PI effective action $\mathbf{\G}[G]$ evaluated on shell is the free energy in the absence of external sources, and we denote it $\mathbf{F}$. In order to test the stability of the conformal solution, we need to consider the effective action at quadratic order in the fluctuations $\d G= G-G_\star$. The expansion at quadratic order can be written schematically as $\mathbf{\G}[G] \simeq \mathbf{F} + \f12 \d G_{12} \cdot G_\star^{-1}{}_{11'} G_\star^{-1}{}_{22'}  (1-K[G_\star])_{1'2'34} \cdot \d G_{34}$, where $K[G_\star]$ is the Bethe-Salpeter kernel and the subscripts stand for the points $x_1$, $x_2$, etc, which are integrated in $\mathbb{R}^d$.

\item The fluctuations $\d G(x,y)$ form a Hilbert space of bilocal functions, for which it is known that a complete and orthonormal basis is provided by a set of functions with the conformal structure of three-point functions $V(x,y;z,h,J) \sim \Braket{\phi(x) \phi(y) \cO_{h}^{\m_1\cdots \m_J}(z)}_{\rm cs}$, with the conformal dimension of the (at this stage unphysical) operator $\cO_{h}$ in the principal series: $h=\f{d}{2}+\im r$. 
Such functions form also an eigenbasis of the Bethe-Salpeter kernel, with eigenvalues $k(h,J)$, parametrized by the conformal dimension $h$ and the spin $J$.
With the aid of such basis, we can therefore write
\be \label{eq:Gamma-quadr-sketch}
\mathbf{\G}[G] - \mathbf{F} \simeq   \, \f18  \sum_{J\in \mathbb{N}_0}   \int_{\frac{d}{2}-\im\infty}^{\frac{d}{2}+\im\infty} \frac{{\rm d}h}{2\pi \im}
   \, \r(h,J)\, (1-k(h,J))    \int \dd  z \, \overline{F}_{h}^{\m_1\cdots \m_{J}}(z)  F_{h}^{\m_1\cdots \m_{J}}(z)  \,,
\ee
where $\r(h)$ is the Plancherel weight, and $F_{h}^{\m_1\cdots \m_{J}}(z)$ are the expansion coefficients of $\d G$ on the basis of $V$-functions.
This and the next part of the proof rely heavily on the  work of Dobrev et al.\cite{Dobrev:1976vr,Dobrev:1975ru,Dobrev:1977qv}, as well as on the more recent literature on conformal partial waves \cite{Caron-Huot:2017vep,Simmons-Duffin:2017nub,Liu:2018jhs,Karateev:2018oml}.\footnote{These methods have been at the heart of a very active field in recent years, see for example their use with Mellin amplitudes  \cite{Mack:2009mi,Costa:2012cb}, their application to the Sachdev-Ye-Kitaev (SYK) model \cite{Maldacena:2016hyu,Murugan:2017eto}, to the bootstrap crossing equations \cite{Gadde:2017sjg,Hogervorst:2017sfd,Sleight:2018ryu,Sleight:2018epi}, and to the construction of an AdS/CFT map \cite{deMelloKoch:2018ivk,Aharony:2020omh}.}

\item The eigenvalues of the Bethe-Salpeter kernel enter also in the conformal partial wave representation of the four-point function, which in fact is essentially the inverse of the Hessian of $\mathbf{\G}[G]$. And in particular, the equation $k(h,J)=1$ defines the poles that lead to the OPE representation. The second main hypothesis of Claim~\ref{claim1}  is then stated as the fact that the equation $k(h,J)=1$, at some fixed $J$, admits a simple root $h_\star$ on the principal series.
Therefore, if $r_\star\neq 0$, the integrand in \eqref{eq:Gamma-quadr-sketch} is negative for ${\rm Im}(h)$ either just above or just below $ r_\star$, and thus there is an instability. For example, if $1-k(h,J)<0$ for ${\rm Im}(h)< r_\star$, choosing $F_{h}^{\m_1\cdots \m_{J}}(z)  $ such that the integral of its squared modulus over $z$ is peaked around $h_\star-\im \eps$, for some $\eps>0$, leads to a negative quadratic fluctuation of the effective action, that is, an instability.

\end{enumerate}

The detailed proof is given in Section~\ref{sec:proof}, and it is written for a generic multiscalar theory; therefore, with respect to what sketched above the formulas result to be complicated by field indices.
The latter could be seen as an unnecessary annoyance from the point of view of the core argument, but they are the price to pay for aiming at a more generic theory than that of a single scalar or even of multiscalars with a large symmetry group (e.g.\ the $O(N)$ model). 

As anticipated, another aspect in which the proof is rather general, is that it does not require a large-$N$ limit. However, in perturbation theory at finite $N$, the Bethe-Salpeter kernel is in general given by an infinite series of diagrams which we do not know to all orders, and whose sum is expected to diverge; this might be only a practical (although crucial) limitation of the perturbative expansion, as we would expect the kernel to exist non-perturbatively in a CFT.  In fact, the Bethe-Salpeter kernel played a central role in the ``old bootstrap" approach to CFTs \cite{Polyakov:1970xd,Migdal:1972tk,Parisi:1972zm,Mack:1972kq,Mack:1973mq}.
And given a conformal four-point function, whose existence is of course implicit in any approach to CFT, one should be able to reconstruct the kernel eigenvalues by inverting the conformal partial wave representation of the correlator \cite{Caron-Huot:2017vep,Simmons-Duffin:2017nub}.
On the other hand, at large-$N$, we are often able to write the Bethe-Salpeter kernel and its eigenvalues in a closed form, thus allowing a direct evaluation of its eigenvalues.

In Section~\ref{sec:examples} we discuss various examples that are made very explicit by the large-$N$ limit. In particular, we consider various models dominated by melonic diagrams, and also the biscalar fishnet model. They mostly serve as a sanity check, as well as a concrete application of our main result.

In Section~\ref{sec:Outlook} we give an outlook on possible generalization, we comment on a conjecture regarding the true vacuum of the models presenting the type of instability discussed here \cite{Kim:2019upg}, and we discuss a simple corollary of the main Proposition on the distinction between real and complex CFTs.

%%%%%%%%%%%%%%%%%%%%%%%%%%%%%
\section{Proof of the main Proposition}
\label{sec:proof}
%%%%%%%%%%%%%%%%%%%%%%%%%%%%%

We consider a field theory formulated as a functional integral over $N$ fundamental fields $\phi_a(x)$, where $x\in\mathbb{R}^d$, and $a=1,\ldots,N$ labels the different fields,\footnote{No $O(N)$ symmetry is assumed. Depending on the specific model, the label could actually represent a collection of labels. For example, it could include both flavor and color indices, e.g.\ for $D$ fields in the bifundamental representation of $O(M)^2$ we would have $a=(A,i,j)$ with $A=1\ldots D$ and $i,j=1\ldots M$ (and hence $N=DM^2$).}  which we assume to be real scalars.\footnote{\label{foot:complexFields}The extension to complex or Grassmannian fields is straightforward. In particular complex fields can be written in terms of real and imaginary parts, so they are included in our treatment (although this is typically not the most convenient way to treat them). Grassmannian fields could be included by assigning a grading to a field label. For fields of spin greater than zero things get significantly more involved, as one should appropriately extend the basis of conformal partial waves used below, because three-point functions of spinning operators admit more than one conformal structure, and because in odd dimensions discrete series representation should be included as well.} 
In the following, we will use a compact notation for the field location and label, thus writing $\phi_a(x)=\phi(X)$ with $X=(x,a)$; accordingly, we will use shorthands such as $\int_X = \sum_a \int \dd x$ and $\d(X-X')=\d_{aa'}\d(x-x')$.
We do not need to specify the bare action $S[\phi]$, we only assume that it is renormalizable; functionals and correlation functions below will be thought to be expressed in terms of renormalized quantities.

%%%%%%%%%%%%%%%%%%%%%%
\paragraph{The 2PI effective action and fluctuations around the conformal solution.}
%%%%%%%%%%%%%%%%%%%%%%
From the functional integral we construct the 2PI effective action  $\mathbf{\G}[G]$ depending only on the bilocal field $G(X,Y)$.\footnote{We are not interested in a possible phase with non-trivial vacuum expectation value for $\phi$, hence we do not introduce a mean field. We should however introduce it if we wish to use the effective action also to generate $n$-point functions with odd $n$.} 
We refer to \cite{Cornwall:1974vz,Berges:2004yj,Benedetti:2018goh} for more details on the 2PI formalism. 
As a brief reminder, the 2PI effective action is constructed by adding  to the action a bilocal source term, $\f12 \int_{X,Y} \phi(X) \cJ(X,Y)\phi(Y)$, and then taking a Legendre transform of the logarithm of the partition function,
\be \label{eq:Gamma-Legendre}
\mathbf{\G}[G]=\left(- {\bf W}[\cJ] +\f12 \Tr[\cJ G]\right){\Big|_{\f{\d {\bf W}}{\d \cJ}=\f12 G}}\,, 
\ee
where 
\be \label{eq:W[J]}
{\bf W}[\cJ]=\ln Z[\cJ] = \ln \int [d\phi] \exp\left\{ -S[\phi] + \f12 \int_{X,Y} \phi(X) \cJ(X,Y)\phi(Y)  \right\} \,.
\ee
In practice, this amounts to computing the free energy only including skeleton (i.e.\ 2PI) diagrams in its perturbative expansion, with propagator $G$, to be determined a posteriori by a self-consistency equation. It thus represents a straightforward generalization of the  usual effective action, that amounts to computing the free energy only including  1PI diagrams, but with a background field, and determining the one-point function from its field equations.
The Schwinger-Dyson (SD) equations for the two-point function are in fact the field equations of the 2PI effective action,
\be
\f{\d \mathbf{\G}}{\d G(X_1,X_2)}\Big|_{{G=G_\star}} = 0\,;
\ee
 that is, their solution, which we denote $G_\star(X,Y)$, is the two-point function of the theory without external sources. 
For a theory obtained as a perturbation of a (generalized) free theory of covariance $C(X,X')$,  the 2PI effective action takes the form
\be \label{eq:Gamma_2PI}
\mathbf{\G}[G] = \f12 \Tr[C^{-1} G]  + \f12 \Tr[\ln G^{-1}] +\mathbf{\G}_2 [G] \,,
\ee
leading to the standard SD equations
\be \label{eq:SDeq-2pt}
G^{-1}(X,X') = C^{-1}(X,X') - \Sigma(X,X') \,,
\ee
with the self energy given by $\Sigma[G] = -2\, \d\mathbf{\G}_2/\d G$. A standard analysis shows that $\mathbf{\G}_2 [G]$ is constructed as a sum of (two- or higher-loop)  2PI diagrams constructed from the vertices of $S[\phi]$, but with $G$ as propagator.
The effective action is thus far written formally, as divergences and renormalization can only be discussed after going on-shell.
The on-shell effective action coincides with the free energy in the absence of external sources: $\mathbf{F}=\mathbf{\G}[G_\star]$; this is of course IR divergent on a non-compact space and should be regulated by working at finite volume, but we will only be interested in the fluctuations, which are not affected by such IR divergence.
Taking two or more derivatives and evaluating them on-shell, one obtains kernels out of which $2n$-point functions can be built, as we will review below for the four-point function. Such on-shell correlators have UV divergences that in a renormalizable theory are taken care of by writing bare couplings and field normalizations as divergent power series in the renormalized couplings; therefore, the bare couplings appearing in \eqref{eq:Gamma_2PI} are determined a posteriori as a function of the solution of \eqref{eq:SDeq-2pt}. The renormalization procedure leads as usual to a renormalization group flow of the renormalized couplings, and scale or conformal invariance of the theory can only be attained at fixed points.

The first main hypothesis of Claim~\ref{claim1} is then stated as follows: 
\begin{hypothesis} \label{hyp1}
Let a Euclidean quantum field theory of $N$ real scalar fields in $\mathbb{R}^d$ be given, and assume that the Schwinger-Dyson equations for the two-point functions, for some choice of renormalized couplings corresponding to a fixed point of the renormalization group, admit a conformal solution\footnote{\label{foot:limitSD}In practice, the solution to the SD equations is in general only found in an asymptotic (e.g.\ IR) limit. All the CFT arguments below will apply to the correlators found by such implicitly assumed (asymptotic and renormalization) limit procedures.}
$$G_\star(X_1,X_2) \sim \d_{a_1 a_2} |x_1-x_2|^{-2\D_1}\,,$$ 
where $\D_i\in\mathbb{R}$ is the scaling dimension of $\phi_{a_i}$; 
moreover, also the four-point functions (and possibly all the other $n$-point functions, the ones with even $n$ being related to functional derivatives of $\mathbf{\G}[G]$ with respect to $G$, evaluated at $G_\star$) are conformal.
\end{hypothesis}

The advantage of phrasing the problem in the 2PI formalism is that we can study the effect of small variations of the two-point function on the free energy.
In order to test the stability of the $G_\star$ solution (or ``vacuum'') we need to consider the effective action at quadratic order in the fluctuations $\d G= G-G_\star$, 
\be \label{eq:Gamma-quadratic}
\mathbf{\G}[G]-  \mathbf{F} \simeq  \f12 \int_{X_1\ldots X_4} \d G(X_1,X_2)  \f{\d^2 \mathbf{\G}}{\d G(X_1,X_2) \d G(X_3,X_4)}\Big|_{{G=G_\star}}  \d G (X_3,X_4) \,.
\ee
and check whether there are perturbations giving a negative contribution. In order to do that, we rely on the \emph{conformal partial wave} expansion of bilocal and four-point functions, as established in \cite{Dobrev:1976vr,Dobrev:1975ru,Dobrev:1977qv}, and which we now review briefly; the reader familiar with such formalism can fast forward to \eqref{eq:f-transform} or beyond.

%%%%%%%%%%%%%%%%%%%%%%
\paragraph{A basis of bilocal functions.}
%%%%%%%%%%%%%%%%%%%%%%
First, by a finite field strength renormalization and a rotation in the space of fields we choose the following normalization of the two-point function:\footnote{The rotation in the space of fields is needed in general for the diagonalization of the two-point function in the subspace of fields having the same dimension. The fact that the latter is diagonalizable follows from excluding the case of a logarithmic CFT.}
\be \label{eq:G_star}
G_\star(X_1,X_2) = \la \phi(X_1) \phi(X_2) \ra = \f{n(\D_1,0) 2^{\D_1}}{(2\pi)^{d/2} } \f{\d_{a_1 a_2} }{|x_{12}|^{2\D_1}} \,,
\ee
with $x_{ij}=x_i-x_j$, and
\be
n(\D,J) = \f{ \G(d-\D-1) \G(\D+J) }{ \G(\f{d}{2}-\D) \G(d-\D-1+J) } \,,
\ee
chosen in such a way that
\be
G_\star^{-1}(X_1,X_2) = G_\star(\wtX_1,\wtX_2) \equiv G_\star(X_1,X_2)\big|_{\D_i \to \widetilde{\D}_i=d-\D_i} \,.
\ee
That is, the inverse two-point function coincides with the two-point function of the shadow operators $\widetilde{\phi}(X_i)$ \cite{Ferrara:1972uq}, which by definition have dimension $\widetilde{\D}_i=d-\D_i$, and which we equivalently denote by placing the tilde on the argument, $\phi(\wtX_i)\equiv \widetilde{\phi}(X_i)$. 
Therefore, the amputation of an external leg in a correlator amounts to performing a shadow transform:
\be
\int_X  G_\star^{-1}(X_1,X)  \la \phi(X) \phi(X_2) \cdots \phi(X_n)  \ra  \propto \la \widetilde{\phi}(X_1) \phi(X_2) \cdots \phi(X_n)  \ra \,.
\ee

We also introduce the two- and three-point functions involving a generic operator $\cO_h^{\m_1\cdots \m_J}$ of dimension $h$ and spin $J$ ($\m_i=1\ldots d$ are $SO(d)$ indices), not necessarily in the physical spectrum of the theory. 
Actually we will mostly be interested in unitary irreducible representations in the  principal series
\be \label{eq:principal-series}
\cP = \left\{ h \Bigm| h=\f{d}{2}+\im r,\, r\in\mathbb{R} \right\}\,,
\ee
and we also define its upper branch $\cP_+$ by the additional constraint $r>0$.

For the two-point function of such operator we choose a normalization that leads to similar properties as for the fundamental fields:\footnote{Repeated spin indices are summed over, and since we are in an Euclidean setting we make no difference between upper and lower indices.}
\be
\begin{split}
G_{h}^{\m_1\cdots \m_J,\n_1\cdots \n_J}(x_1,x_2) & =   \f{n(h,J) 2^{h}}{(2\pi)^{d/2} } \la \cO_h^{\m_1\cdots \m_J}(x_1)  \cO_h^{\n_1\cdots \n_J}(x_2) \ra_{\rm cs} \\
& = \f{n(h,J) 2^{h}}{(2\pi)^{d/2} } \f{ \cS^{\m_1\cdots \m_J}_{\m'_1\cdots \m'_J}\, \cS^{\n_1\cdots \n_J}_{\n'_1\cdots \n'_J}\, I^{\m'_1\n'_1} \cdots I^{\m'_J\n'_J}}{|x_{12}|^{2h}} \,,
\end{split}
\ee
where the subscript cs stands for ``conformal structure" (with standard normalization), $\cS^{\m_1\cdots \m_J}_{\m'_1\cdots \m'_J}$ is the projector on the symmetric-traceless part of rank-$J$ tensors, and
\be
I^{\m\n}= \d^{\m\n} -  2 \f{x_{12}^\m x_{12}^\n}{|x_{12}|^2} \,.
\ee
Also in this case the inverse two-point function coincides with the two-point function of the shadow operator $\cO_{\htilde}$.

Next, we introduce the following vertex functions, or $V$-functions, having the conformal structure of three-point functions of one $\cO_h$ and two $\phi$:
\be \label{eq:V-def}
\begin{split}
V_{h;\s}^{\m_1\cdots \m_J}(X_1,X_2; x_3) 
&= \cN^{\D_1,\D_2}_{h,J} \la \phi_{\D_1}(x_1) \phi_{\D_2}(x_2) \cO_h^{\m_1\cdots \m_J}(x_3) \ra_{\rm cs}\, E^{\s,J}_{a_1 a_2} \\ 
&= \cN^{\D_1,\D_2}_{h,J} \f{ \cS^{\m_1\cdots \m_J}_{\n_1\cdots \n_J} \, Z^{\n_1} \cdots Z^{\n_J} }{ |x_{12}|^{\D_1+\D_2-h+J} |x_{13}|^{\D_1-\D_2+h-J} |x_{23}|^{\D_2-\D_1+h-J} } E^{\s,J}_{a_1 a_2}
\,,
\end{split}
\ee
where 
\be
Z^{\m} = \f{x_{13}^{\m}}{|x_{13}|^2} - \f{x_{23}^{\m}}{|x_{23}|^2} \,.
\ee
In the first line of \eqref{eq:V-def} we have used the field dimensions as subscript for $\phi$, as the conformal structure depends only on that, and not on the field label. 

The matrix structure $E^{\s,J}_{a_1 a_2}$ is at this stage rather free, and it could depend parametrically on the spin and conformal dimensions. We have only marked explicit the dependence on spin, as for sure it has to depend on its parity, in the following way: the set of matrices $E^{\s,J}_{a_1 a_2}$, with $\s=1\ldots N(N+(-1)^J)/2$, should form an orthonormal complete basis in the space of real symmetric or antisymmetric $N\times N$ matrices,  for $J$ even or odd, respectively.
That is,  they satisfy the orthonormality
\be
\sum_{a_1,a_2} E^{\s,J}_{a_1 a_2} E^{\s',J}_{a_1 a_2} = \d_{\s\s'} \,,
\ee
and completeness relation
\be
\sum_{\s} E^{\s,J}_{a_1 a_2} E^{\s,J}_{a_3 a_4} =
\f12 (\d_{a_1 a_3} \d_{a_2 a_4} + (-1)^J \d_{a_1 a_4} \d_{a_2 a_3} ) \,.
\ee
The rationale behind the symmetry requirement on the matrix elements is that ultimately the $V$-functions will serve as basis for objects with the same symmetries as the product $\phi(X_1)\times\phi(X_2)\equiv \phi_{a_1}(x_1)\times \phi_{a_2}(x_2)$, which is symmetric under the exchange $X_1\leftrightarrow X_2$, but in general not under $x_1\leftrightarrow x_2$ alone.
Since $\la \phi_{\D_1}(x_1) \phi_{\D_2}(x_2) \cO_h^{\m_1\cdots \m_J}(x_3) \ra_{\rm cs}$ is symmetric or antisymmetric  under  $(x_1,\D_1)\leftrightarrow (x_2,\D_2)$ for even or odd spin, respectively, the symmetry properties of $E^{\s,J}_{a_1 a_2}$ lead to a $V_{h;\s}^{\m_1\cdots \m_J}(X_1,X_2; x_3)$ which is symmetric in $X_1\leftrightarrow X_2$. 
A further specification of the matrix basis will be needed later when diagonalizing the Bether-Salpeter kernel.
.

The normalization factor is chosen to be\footnote{See \cite{Dobrev:1976vr} for the choice of branch of the square root.}
\be
\begin{split}
\cN^{\D_1,\D_2}_{h,J} = & \f{2^{(\D_1+\D_2+h+J)/2}}{(2\pi)^{d/2}} \\
&\quad\times\left(  \frac{\G(\tfrac{\htilde+J+\D_1+\D_2-d}{2}) \G(\tfrac{h+J+\D_1+\D_2-d}{2}) \G(\tfrac{h+J +\D_1-\D_2}{2}) \G(\tfrac{h+J-\D_1+\D_2}{2}) }{ \G(\tfrac{\htilde+J-\D_1-\D_2+d}{2}) \G(\tfrac{h+J-\D_1-\D_2+d}{2}) \G(\tfrac{\htilde+J +\D_1-\D_2}{2}) \G(\tfrac{\htilde+J-\D_1+\D_2}{2})  }   \right)^{1/2} \,,
\end{split}
\ee
such that:
\begin{itemize}
\item the shadow transform of $V$ has unit coefficient, i.e.\ under amputation of external legs we have:
\be \label{eq:V-shadow}
\int_X  G_\star^{-1}(X_1,X) V_{h;\s}^{\m_1\cdots \m_J}(X,X_2; x_3) = V_{h;\s}^{\m_1\cdots \m_J}(\wtX_1,X_2; x_3) \,,
\ee
together with a similar formula on the second argument, and
\be \label{eq:V-shadow-h}
\int \dd  z \, G_{\htilde}^{\m_1\cdots \m_J,\n_1\cdots \n_J}(x_3,z)V_{h;\s}^{\n_1\cdots \n_J}(X_1,X_2; z) = V_{\htilde;\s}^{\m_1\cdots \m_J}(X_1,X_2; x_3) \,.
\ee
\item For $h\in\cP$, and $\D_1$ and $\D_2$ satisfying the condition\footnote{See ``Conclusion 10.8'' of \cite{Dobrev:1977qv}. The range of applicability can however be extended by analytic continuation, if we appropriately deform the integration contour, see for example the discussion in \cite{Aharony:2020omh}.}
\be \label{eq:Deltas-conditions}
|\mathrm{Re}(\D_1-\f{d}{2})|+|\mathrm{Re}(\D_2-\f{d}{2})|\leq \f{d}{2} \,,
\ee
the following orthonormalization property holds:
\be \label{eq:V-orthonorm}
\begin{split}
\int_{X_1,X_2}  &V_{\htilde;\s}^{\m_1\cdots \m_J} (\wtX_1,\wtX_2; z) V_{h';\s'}^{\n_1\cdots \n_{J'}}(X_1,X_2; z') \\
& = \f{2\pi\im}{\r(h',J)} \d_{J'J} \d_{\s\s'}\\
&\qquad \times\left( \d(z-z') \d(h-h') \cS^{\m_1\cdots \m_J}_{\n_1\cdots \n_{J}}  + G_{h'}^{\m_1\cdots \m_J,\n_1\cdots \n_J}(z-z') \d(\htilde-h') \right)  \,,
\end{split}
\ee
where we have introduced the Plancherel weight
\be
\begin{split}
\r(h,J) &= \f{\G(\tfrac{d}{2}+J)}{2 (2\pi)^{d/2} J!} \f{\G(\htilde-1)\G(h-1)}{\G(\f{d}{2}-h)\G(\f{d}{2}-\htilde)} (h+J-1) (\htilde+J-1) \\
&= \f{\G(\tfrac{d}{2}+J)}{2 (2\pi)^{d/2} J!} n(h,J) n(\htilde,J) \,.
\end{split}
\ee

\item Under complex conjugation, for real $\D_1$ and $\D_2$, and $h\in\cP$, we have
\be \label{eq:V-conj}
\overline{V_{h;\s}^{\m_1\cdots \m_J} (X_1,X_2; x_3)}%^*
 = V_{\htilde;\s}^{\m_1\cdots \m_J}(X_1,X_2; x_3) \,.
\ee

\item For $d>1$, and under condition \eqref{eq:Deltas-conditions}, the following completeness relation holds:\footnote{This is the right-amputated version of the conformal partial wave representation of the $s$-channel-connected four-point function of a generalized free field theory, or mean field theory, i.e.\ $G_\star(X_1,X_3) G_\star(X_2,X_4)+G_\star(X_1,X_4) G_\star(X_2,X_3)$ \cite{Fitzpatrick:2011dm,Liu:2018jhs,Karateev:2018oml}.}
\be \label{eq:V-complet}
\begin{split}
\sum_{J\in \mathbb{N}_0}  \int \dd  z & \int_{\cP_+}%{\f{d}{2}}^{\f{d}{2}+\im\infty}
  \f{{\rm d}h}{2\pi\im} \r(h,J) \sum_{\s} V_{h;\s}^{\m_1\cdots \m_{J}}(X_1,X_2; z) V_{\htilde;\s}^{\m_1\cdots \m_{J}}(\wtX_3,\wtX_4; z) \\
&= \f12\left( \d(X_1-X_3)\d(X_2-X_4) + \d(X_1-X_4)\d(X_2-X_3) \right) \\
&\equiv  \mathbb{I}(X_1,X_2,X_3,X_4)\,,
\end{split}
\ee
where $\mathbb{I}$ is the identity operator in the space of symmetric bilocal functions.
For comparison to the litterature, it is worth noticing that the identity operator can also be written as
\be
\begin{split}
\mathbb{I}( &X_1,X_2,X_3,X_4) =\\
& \f12 (\d(x_1-x_3) \d(x_2-x_4)+\d(x_1-x_4)\d(x_2-x_3))\f12 (\d_{a_1 a_3} \d_{a_2 a_4} +  \d_{a_1 a_4} \d_{a_2 a_3} ) \\
&+ \f12 (\d(x_1-x_3) \d(x_2-x_4)-\d(x_1-x_4)\d(x_2-x_3))\f12 (\d_{a_1 a_3} \d_{a_2 a_4} -  \d_{a_1 a_4} \d_{a_2 a_3} ) \,,
\end{split}
\ee
where the first line comes from the contribution of even $J$ in \eqref{eq:V-complet}, and the second from odd $J$.
If $a_1=a_2$ (and hence $a_3=a_4$), the odd $J$ contributions vanish (as expected, since $E^{\s,J}_{a a}=0$).
If instead $a_1=a_3 \neq a_2 = a_4$, then the whole expression reduces to just $\f12 \d(x_1-x_3) \d(x_2-x_4)$.

For $d=1$, the discrete series representations $h\in \{h=2n \bigm| n\in \mathbb{N}\}$ need to be included as well \cite{Maldacena:2016hyu}.\footnote{The discrete series representations exist for any odd $d$, however they do not appear in the decomposition of two scalar representations for $d>1$ (Proposition~2.1 of \cite{Dobrev:1976vr}, or Theorem~10.5 of \cite{Dobrev:1977qv}).}
From now on we assume $d>1$, and we will comment at the end about the $d=1$ case.

\end{itemize}

Notice that in the orthonormality relation the second term on the right-hand side vanishes if we restrict both $h$ and $h'$ to be in $\cP_+$.
Therefore, interpreting the location of $\cO_h$ in\eqref{eq:V-def}, together with its dimension $h$, spin, and relative indices as labels of a set of bilocal functions (of $X_1$ and $X_2$), the equations \eqref{eq:V-orthonorm} and \eqref{eq:V-complet} state their orthonormality and completeness with respect to the inner product
\be \label{eq:scalar-prod}
\begin{split}
(f_1 , f_2) = \f12 \int_{X_1\ldots X_4 } \overline{f_1(X_1,X_2)} & \left( G_\star^{-1}(X_1,X_3) G_\star^{-1}(X_2,X_4) \right. \\
& \quad\left.+G_\star^{-1}(X_1,X_4) G_\star^{-1}(X_2,X_3) \right) f_2(X_3,X_4) \,,
\end{split}
\ee
with $f_i(X_1,X_2)$ in the space $\mathcal{V}$ of smooth symmetric functions which are square integrable with respect to the measure in \eqref{eq:scalar-prod} and satisfy the asymptotic boundary condition $f_i(X_1,X_2) \sim |x_1|^{-2\D_1}$ for $|x_1|\to\infty$ and $f_i(X_1,X_2) \sim |x_2|^{-2\D_2}$ for $|x_2|\to\infty$.\footnote{More precisely, the space of functions is in the union of Kronecker products of two type I (scalar) complementary series representations, satisfying condition \eqref{eq:Deltas-conditions} (see sections 10.B-D of \cite{Dobrev:1977qv}, and in particular ``Conclusion 10.8'').
The $V$-functions are not in $\mathcal{V}$, as they are not integrable, but they form a basis in the continuous sense, just like the Fourier basis does for ${\rm L}^2(\mathbb{R}^d)$.}
It is also useful to introduce the shadow space $\tilde{\mathcal{V}}$ which corresponds to the replacement $\D_i\to\wtD_i$ in $\mathcal{V}$.
As a consequence of the completeness and orthonormality of the $V$-functions, any function $f\in\mathcal{V}$ can be represented as
\be  \label{eq:f-transform}
f(X_1,X_2) =  \sum_{J\in \mathbb{N}_0} \int \dd  z \int_{\cP_+}%{\f{d}{2}  }^{\f{d}{2}+\im\infty}
 \f{{\rm d}h}{2\pi\im} \r(h,J)   \sum_{\s} V_{\htilde;\s}^{\m_1\cdots \m_{J}}(X_1,X_2; z) F_{h;\s}^{\m_1\cdots \m_{J}}(z) \,,
\ee
with coefficients given by the inverse transform
\be \label{eq:F-transform}
F_{h;\s}^{\m_1\cdots \m_{J}}(z) =  \int_{X_1,X_2}   V_{h;\s}^{\m_1\cdots \m_{J}}(\wtX_1,\wtX_2; z) f(X_1,X_2) \,.
\ee
Notice that for $a_1=a_2$, as pointed out above, only even spins contribute.

For real functions, the complex conjugation property \eqref{eq:V-conj}, together with the definition \eqref{eq:F-transform}, implies a similar property for the transform of $f$, namely: 
\be \label{eq:F-conj}
\overline{F_{h;\s}^{\m_1\cdots \m_{J}}(z)}= F_{\htilde;\s}^{\m_1\cdots \m_{J}}(z)\,.
\ee
This is reminiscent of the complex conjugation property of the Fourier transform $F(p)$ of a real function $f(x)$, namely: $\overline{F}(p)=F(-p)$. 
However, in the present case, it is not as straightforward to see that the complex conjugation property of $F$ implies the reality of $f$ in \eqref{eq:f-transform}.
In fact, this is not immediately obvious even in the left-hand side of \eqref{eq:V-complet}; reality follows because such expression is non-zero only when the pair $(X_1,X_2)$ is equal to the pair $(X_3,X_4)$ (up to permutation).
In order to make manifest the reality of \eqref{eq:f-transform}, under the condition \eqref{eq:F-conj}, we can extend the integration to the whole principal series:
\be  \label{eq:f-transform-real}
f(X_1,X_2) = \f12 \sum_{J\in \mathbb{N}_0}  \int \dd  z \int_{\cP}
  \f{{\rm d}h}{2\pi\im} \r(h,J)  \sum_{\s}  V_{\htilde;\s}^{\m_1\cdots \m_{J}}(X_1,X_2; z) F_{h;\s}^{\m_1\cdots \m_{J}}(z) \,.
\ee
In this representation it is the inverse transform which is less obvious, due to the second term in \eqref{eq:V-orthonorm}, and one has to take into account the property
\be \label{eq:F-shadow-h}
\int \dd  z'  G_{h}^{\m_1\cdots \m_J,\n_1\cdots \n_J}(z,z') F_{\htilde;\s}^{\n_1\cdots \n_{J}}(z') = F_{h;\s}^{\m_1\cdots \m_{J}}(z) \,,
\ee
which is consistent with \eqref{eq:F-transform}, because of \eqref{eq:V-shadow-h}.

%%%%%%%%%%%%%%%%%%%%%%
\paragraph{Hessian of $\mathbf{\G}[G]$ and OPE spectrum.}
%%%%%%%%%%%%%%%%%%%%%%
Going back to \eqref{eq:Gamma-quadratic}, and using \eqref{eq:Gamma_2PI}, we see immediately that for a free theory, i.e.\ for $\mathbf{\G}_2 [G]=0$, the part of $\mathbf{\G}[G]$ quadratic in the fluctuations takes the form of the inner product \eqref{eq:scalar-prod}, with $f_1$ and $f_2$ replaced by $\d G$.
Therefore, we naturally have $\d G(X_1,X_2)\in \mathcal{V}$, and we can represent the fluctuation fields as in \eqref{eq:f-transform-real}; we promote this property to the interacting theory.

In the interacting case, we write the Hessian of the 2PI effective action in terms of the Bethe-Salpeter kernel $K$ defined by
\be \label{eq:Hess-K}
\f{\d^2 \mathbf{\G}[G]}{\d G(X_1,X_2) \d G(X_3,X_4)}\Big|_{{G=G_\star}}
   = \f12 \int_{Y_1,Y_2} G_\star^{-1}(X_1,Y_1)G_\star^{-1}(X_2,Y_2) \left(\mathbb{I}-K\right)(Y_1,Y_2,X_3,X_4) \,,
\ee
or, using \eqref{eq:Gamma_2PI},
\be \label{eq:K}
K(X_1,X_2,X_3,X_4) =  -2 \int_{Y_1,Y_2}  G_\star(X_1,Y_1)G_\star(X_2,Y_2) \f{\d^2 \mathbf{\G}_2[G] }{\d G({Y_1,Y_2}) \d G(X_3,X_4)}\Big|_{{G=G_\star}} \,.
\ee
Notice that since $\mathbf{\G}_2[G]$ is a sum of 2PI vacuum diagrams, the Bethe-Salpeter kernel is 2PI in the $s$-channel ($12\to34$), that is, the $x_1$ and $x_2$ external points cannot be disconnected from $x_3$ and $x_4$ by cutting less than three internal lines. Notice also that the Hessian is completely amputated (no propagators on the external legs), while $K$ is only amputed on the right (legs $(3,4)$).

\begin{figure}[ht]
\begin{center}
\includegraphics[width=0.9\textwidth]{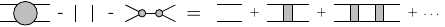}
 \caption{Graphical representation of $\cF_s$, combining \eqref{eq:F_s}, on the left-hand side, with \eqref{eq:GF=1} and \eqref{eq:Hess-K}, on the right-hand side. The grey disks with four and three legs represent the full four-point and three-point functions, respectively, while a grey rectangle, together with the two edges to its left, represents the Bethe-Salpeter kernel. For simplicity we have omitted the crossed graphs with $X_3$ and $X_4$ exchanged.} \label{fig:F-ladder}
 \end{center}
\end{figure}

The four-point objects defined above can be viewed as kernels of linear maps from some space of bilocal functions to another, whose composition translates in a matrix-like multiplication of the kernels, defined by the convolution on their left $(1,2)$ or right $(3,4)$ pair of indices, as for example on the right-hand sides of \eqref{eq:Hess-K} for the product of $G_\star^{-1}G_\star^{-1}$ with $(\mathbb{I}-K)$.
A standard Bethe-Salpeter-like equation (e.g.\ \cite{Dobrev:1975ru}) then implies that, with respect to such multiplication, the Hessian \eqref{eq:Hess-K} is the inverse of the four-point function connected and 1PI in the $s$-channel, that is:
\be \label{eq:GF=1}
\int_{Y_1,Y_2} \f{\d^2 \mathbf{\G}[G]}{\d G(X_1,X_2) \d G(Y_1,Y_2)}\Big|_{{G=G_\star}} \cF_s(Y_1,Y_2,X_3,X_4) = \mathbb{I}(X_1,X_2,X_3,X_4)\,,
\ee
with
\be \label{eq:F_s}
\begin{split}
\cF_s(X_1,X_2,X_3,X_4)   \equiv & \langle{\phi(X_1) \phi(X_2) \phi(X_3) \phi(X_4)} \rangle 
 -   G_\star(X_1,X_2)  G_\star(X_3,X_4) 
\\
 &-\int_{Y_1,Y_2} \langle \phi(X_1) \phi(X_2) \ph(Y_1) \rangle G_\star^{-1}(Y_1,Y_2) \langle \phi(Y_2) \phi(X_3) \phi(X_4) \rangle \,.
\end{split}
\ee
For a proof of this relation in the case of vanishing three-point functions $\la \phi \phi \phi\ra$  see \cite{Benedetti:2018goh}. The general case can be proved diagrammatically (essentially noticing that since $\cF_s$ is 1PI in the $s$-channel, it can be written as a sum of ladders whose rungs are 2PI in the $s$-channel, i.e.\ as a geometric series in the Bethe-Salpeter kernel, see Figure~\ref{fig:F-ladder}), or by the type of relations in footnote 11 of \cite{Cornwall:1974vz}.

\
%%%%%%%%%%%%%%
%%%%%%%%%%%%%%

The hypothesis of conformal invariance and the fact that the right legs of $K$ are amputated, imply that the Bethe-Salpeter kernel transforms in the $\D_1\times \D_2\times \wtD_3\times \wtD_4$ representation of the conformal group, and hence it defines an endomorphism on $\mathcal{V}$.
Similarly, the Hessian is a morphism from  $\mathcal{V}$ to  $\tilde{\mathcal{V}}$, while $\cF_s$ maps $\tilde{\mathcal{V}}$ to $\mathcal{V}$.
As the amputation or attachment of external legs is easily taken care of, when we want to obtain a non-perturbative expression for $\cF_s$ from the 2PI effective action, the non-trivial part is the inversion of $\mathbb{I}-K$. Being the latter an endomorphism on $\mathcal{V}$, the natural strategy is to diagonalize it first. To that end, one notices that since the conformal structure of a three-point function involving two scalars is unique, when $K$ acts on a three-point function $\la \phi_{\D_3}(x_3) \phi_{\D_4}(x_4) \cO_h^{\m_1\cdots \m_J}(z) \ra_{\rm cs}$ it must produce a result proportional to $\la \phi_{\D_1}(x_1) \phi_{\D_2}(x_2) \cO_h^{\m_1\cdots \m_J}(z) \ra_{\rm cs}$. 
When acting on a $V$-function we thus obtain
\be
\begin{split}
\int_{X_3,X_4} & K(X_1,X_2,X_3,X_4) V_{h;\s}^{\m_1\cdots \m_J}(X_3,X_4; z) \\
&\quad  = \sum_{a_3 a_4} \hat{k}_{a_1 a_2, a_3 a_4}(h,J) E^{\s,J}_{a_3 a_4} \cN^{\D_1,\D_2}_{h,J} \la \phi_{\D_1}(x_1) \phi_{\D_2}(x_2) \cO_h^{\m_1\cdots \m_J}(z) \ra_{\rm cs}\,,
\end{split}
\ee
where we have absorbed a factor $\cN^{\D_3,\D_4}_{h,J}/\cN^{\D_1,\D_2}_{h,J}$ into the definition of $\hat{k}_{a_1 a_2, a_3 a_4}(h,J)$.

With respect to the inner product \eqref{eq:scalar-prod}, the adjoint operator $K^\dagger$, defined by the identity $(f_1,Kf_2)=(K^\dagger f_1,f_2)$, is
\be
\begin{split}
K^\dagger(X_1,X_2, X_3,X_4) &=  \int_{Y_1\ldots Y_4} \f12 \left( G_\star^{-1}(X_3,Y_3) G_\star^{-1}(X_4,Y_4) +G_\star^{-1}(X_3,Y_4) G_\star^{-1}(X_4,Y_3) \right)\\
& \qquad\times \overline{K(Y_3,Y_4,Y_1,Y_2)} \\
& \qquad\quad \times  \f12 \left( G_\star(Y_1,X_1) G_\star(Y_2,X_2) +G_\star(Y_1,X_2) G_\star(Y_2,X_1) \right)\\
&= \overline{K(X_1,X_2,X_3,X_4)} \,,
\end{split}
\ee
where in the last line we used the definition \eqref{eq:K}, together with the symmetry of the Hessian.
Therefore, if the kernel is real, as we will assume, then it is also self-adjoint, and thus it is diagonalizable.  
From the same argument also follows that $\hat{k}_{a_1 a_2, a_3 a_4}(h,J)$ with $h\in\cP$ is a real symmetric matrix, hence it is diagonalizable, with real eigenvalues, and we can choose the matrices $E^{\s,J}_{a_1 a_2}$ in \eqref{eq:V-def} to be its basis of eigenvectors, i.e.\ we can take them to satisfy
\be
\sum_{a_3 a_4} \hat{k}_{a_1 a_2, a_3 a_4}(h,J) E^{\s,J}_{a_3 a_4} =  k_\s(h,J) E^{\s,J}_{a_1 a_2}\,,
\ee
or equivalently
\be \label{eq:eigV}
\int_{X_3,X_4}  K(X_1,X_2,X_3,X_4) V_{h;\s}^{\m_1\cdots \m_J}(X_3,X_4; z) = k_\s(h,J)V_{h;\s}^{\m_1\cdots \m_J}(X_1,X_2; z) \,.
\ee
Notice that the eigenvalue equation, together with the reality of the kernel and equation \eqref{eq:V-conj}, implies that $\overline{k_{\s}(h,J)}=k_{\s}(\htilde,J)$, for $h\in\cP$.
And since the eigenvalue must also be real, we conclude that $k_{\s}(\htilde,J)=k_{\s}(h,J)$, which by analytic continuation holds for any $h$ in the domain of analyticity.

We can now express the endomorphism $\cF_s$ in the basis of eigenfunctions of $K$, which by the choice of $E^{\s}_{a_1 a_2}$ above is identified with the $V$-function basis of $\mathcal{V}$.
The result is nothing but the conformal partial wave representation of $\cF_s$, i.e.:
\begin{equation} \label{eq:CPW-rep}
\begin{split}
\cF_s(X_1,X_2,X_3,X_4)  
&   =  \sum_{J\in \mathbb{N}_0}   \int_{\cP_+}
 \frac{{\rm d}h}{2\pi \im}  \sum_{\s}
   \, \frac{2\, \r(h,J)}{1-k_{\s}(h,J)} \\ 
   &\qquad\qquad\quad \times  \int \dd  z  \, V_{h;\s}^{\m_1\cdots \m_{J}}(X_1,X_2; z) V_{\htilde;\s}^{\m_1\cdots \m_{J}}(X_3,X_4; z)  \\
&   =  \sum_{J\in \mathbb{N}_0}   \int_{\cP_+}
 \frac{{\rm d}h}{2\pi \im}  \sum_{\s}
   \, \frac{2\, \r(h,J)}{1-k_{\s}(h,J)} \\ 
   &\qquad\qquad\quad \times \, \cN^{\D_1,\D_2}_{h,J}  \cN^{\D_3,\D_4}_{\htilde,J} \,
     \Psi_{h,J}^{\D_i}(x_i) E^{\s,J}_{a_1 a_2}E^{\s,J}_{a_3 a_4} \\
&   =    \sum_{J\in \mathbb{N}_0} 
  \int_{\cP}
   \frac{{\rm d}h}{2\pi \im}  \sum_{\s}
   \,  \frac{2\, \hat{\r}_{\D_i}(h,J)}{1-k_{\s}(h,J)} 
     \, \cG_{h,J}^{\D_i}(x_i) E^{\s,J}_{a_1 a_2}E^{\s,J}_{a_3 a_4}\,,
\end{split}
\end{equation}
where in the second equality we introduced the conformal partial wave, defined as\footnote{In the literature the same name has been often attributed to different objects appearing in this formalism, in particular to what we call $\cG_{h,J}^{\D_i}(x_i)$ (e.g.\ \cite{Dolan:2003hv,Dolan:2011dv,SimmonsDuffin:2012uy,Poland:2018epd}), or 
to the $(n-1)$-point function appearing in the expansion of an $n$-point function  in \cite{Dobrev:1975ru} (i.e.\ the $V$-functions themselvesfor $n=4$). We follow \cite{Simmons-Duffin:2017nub,Liu:2018jhs,Karateev:2018oml} in the choice of naming.}
\be \label{eq:CPW}
 \Psi_{h,J}^{\D_i}(x_i) = \int \dd  z \la \phi_{\D_1}(x_1) \phi_{\D_2}(x_2) \cO_h^{\m_1\cdots \m_J}(z) \ra_{\rm cs}\la {\phi}_{\D_3}(x_3){\phi}_{\D_4}(x_4) {\cO}_{\htilde}^{\m_1\cdots \m_J}(z)\ra_{\rm cs} \,.
\ee
In the last line we used the relation  between $\Psi_{h,J}^{\D_i}$ and conformal blocks $\cG_{h,J}^{\D_i}$  \cite{Simmons-Duffin:2017nub},
\be
\Psi_{h,J}^{\D_i}(x_i) =  \left(-\f12\right)^J  S_{\tilde{h},J}^{\D_3,\D_4}\, \cG_{h,J}^{\D_i}(x_i)  +  \left(-\f12\right)^J  S_{h,J}^{\D_1,\D_2}\, \cG_{\tilde{h},J}^{\D_i}(x_i) \,,
\ee
with
\be
S_{h,J}^{\D_1,\D_2} =
\f{\pi^{d/2} }{ n(\htilde,J)} \f{ \G(\f{\tilde{h}+\D_1-\D_2+J}{2})\G(\f{\tilde{h}+\D_2-\D_1+J}{2})}{ \G(\f{h+\D_1-\D_2+J}{2})\G(\f{h+\D_2-\D_1+J}{2})} \,,
\ee
and we have defined the rescaled measure\footnote{As a comparison to previous literature, we notice that restricting to $a_1=a_2$ and considering only fields with the same dimension $\D$, only the even spin contributions survive, and we have
\be \nn
\hat{\r}_{\D}(h,J)= \left(\f{2^{\D}}{(2\pi)^{d/2}} \f{\G(\D)}{\G(\f{d}{2}-\D)} \right)^2  \f12 \m_{\D}(h,J)= \left(\f{2^{\D}}{(2\pi)^{d/2}} n(\D,0) \right)^2  \f12 \m_{\D}(h,J)\,,
\ee
where we introduced $\m_{\D}(h,J)$ from equation (92) of \cite{Benedetti:2019ikb} (or A.5 of \cite{Benedetti:2020iku}), which was derived from the formulas in \cite{Liu:2018jhs,Karateev:2018oml}. The squared prefactor originates from the different normalization we use here for $G_\star(X_1,X_2)$ (see \eqref{eq:G_star}, and remember that for a free theory we have $K=0$ and $\cF_s(X_1,X_2,X_3,X_4)=G_\star(X_1,X_3)G_\star(X_2,X_4)+G_\star(X_1,X_4)G_\star(X_2,X_3)$), while the $1/2$ factor cancels with the factor 2 in \eqref{eq:CPW-rep} (originating from the $1/2$ in \eqref{eq:Hess-K}).} 
\be \label{eq:rho-hat}
\begin{split}
\hat{\r}_{\D_i}(h,J) &=   \left(-\f12\right)^J  \r(h,J) \, \cN^{\D_1,\D_2}_{h,J}  \cN^{\D_3,\D_4}_{\htilde,J}  \, S_{\tilde{h},J}^{\D_3,\D_4}\\
& = \left(-\f12\right)^J  \f{\G(\tfrac{d}{2}+J)}{2^{1+h}  J!}  n(\htilde,J) \, \cN^{\D_1,\D_2}_{h,J}\cN^{\D_3,\D_4}_{h,J}\,.
\end{split}
\ee
As a consistency check of the discussion on the diagonalization, we notice that since the Plancherel weight and the product $\cN^{\D_1,\D_2}_{h,J}  \cN^{\D_3,\D_4}_{\htilde,J} \Psi_{h,J}^{\D_i}(x_i) $ are real
for $\D_i\in \mathbb{R}$ and $h\in\cP$,\footnote{To show that the product $\cN^{\D_1,\D_2}_{h,J}  \cN^{\D_3,\D_4}_{\htilde,J} \Psi_{h,J}^{\D_i}(x_i) $ is real one can use the relation 
\be \nn
\Psi_{\htilde,J}^{\D_i}(x_i) = \f{S_{h,J}^{\D_3,\D_4}}{S_{h,J}^{\D_1,\D_2}}  \Psi_{h,J}^{\D_i}(x_i) \,,
\ee
and the structure of the normalization factors.
} and $E^{\s,J}_{a_1 a_2}$ is real by assumption, the four point function is real if and only if the eigenvalues $k_{\s}(h,J)$ are real as well.

As the conformal block decays exponentially for positive ${\rm Re}(h)$, one can push the contour in \eqref{eq:CPW-rep} to the right in the complex plane. Assuming that  $1/(1-k_{\s}(h,J))$ is a meromorphic function in this half-plane, then one picks up the poles at $k_{\s}(h,J)=1$, and further assuming that these are only simple poles, the  conformal partial wave representation reduces to the standard OPE in the $s$-channel \cite{Dobrev:1975ru,Simmons-Duffin:2017nub,Liu:2018jhs,Karateev:2018oml}.\footnote{Poles from the measure and from the conformal block are spurious and cancel with each other \cite{Simmons-Duffin:2017nub}.} Therefore, the unit eigenvalues of the kernel $K$ determine the spectrum of primary operators appearing in the OPE of two fundamental fields, and the OPE coefficients are provided by the residue of $\hat{\r}(h,J)/(1-k_{\s}(h,J))$. 
Double poles can appear at transitions points, when by varying a parameter of the theory (e.g.\ the dimension $d$, the number of fields, or some marginal coupling) two simple poles merge and then split again, for example moving away from the real axis in the process. We will stay away from such transition points.

The second main hypothesis of Claim~\ref{claim1} is now stated as:
\begin{hypothesis} \label{hyp2}
Let $K(X_1,X_2,X_3,X_4)$ be the Bethe-Salpeter kernel of the conformal field theory of Hypothesis~\ref{hyp1}, and assume that it is real, and hence diagonalizable, with eigenvalue $k_{\s}(h,J)$, which for each $J$ and $\s$ is real on $h\in\cP$, and 
analytically continued to a meromorphic function in the half-plane  ${\rm Re}(h)\geq d/2$.

Moreover, let the equation $k_{\s}(h,J)=1$ admit, for some fixed $J$ and $\s$, a simple root of the form  $h=h_\star\equiv \f{d}{2}+\im  r_\star$, with $ r_\star\in\mathbb{R}$ and different from zero.
\end{hypothesis}

%\

As $h_\star$ belongs to the principal series, i.e.\ it lies on the initial contour, one has to be careful in defining \eqref{eq:CPW-rep}: the contour needs to be slightly deformed such that for example the pole at $\f{d}{2}+\im  r_\star$ is to its right, while its shadow pole at $\f{d}{2}-\im  r_\star$ is to its left.

\begin{figure}[ht]
\begin{center}
\includegraphics[width=0.9\textwidth]{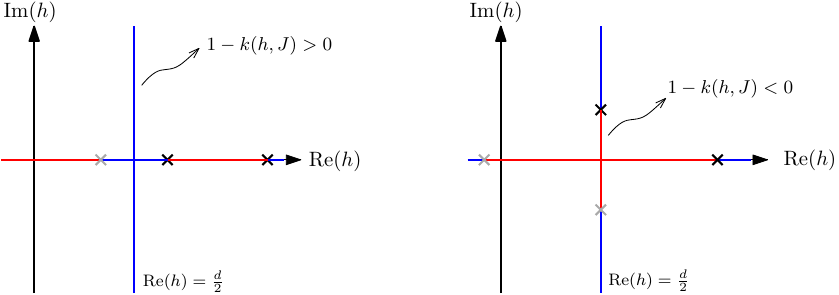}
 \caption{Illustration in the complex $h$ plane of some hypothetical solutions of $k(h,J)=1$. Physical solutions are represented by black crosses, while their shadow by  gray crosses. On the principal series (vertical line) and the real line we have marked in blue the intervals with $1-k(h,J)>0$ and in red those with $1-k(h,J)<0$.
 On the left panel all solutions are real, while on the right panel the smallest physical one has merged with its shadow to form a complex pair, thus leading to a negative interval on the principal series.
 Sign changes can occur also at poles of $k(h,J)$, which we have ignored here for simplicity.} 
 \label{fig:merge-transition}
 \end{center}
\end{figure}

As we will see in the following, such roots typically arise when, by tuning a parameter of the theory, a physical root and its shadow move along the real axis and merge at $h=d/2$, to then split again along the principal series, as depicted in Figure~\ref{fig:merge-transition}. It would therefore look like the transition is precisely of the type described above, giving a double pole at the merging point. However, this case is different from that in which two physical poles merge, as the shadow pole remains unphysical even at the merging. In fact, while $1/(1-k_{\s}(h,J))$ would certainly give a double pole in this case, it turns out that the measure \eqref{eq:rho-hat} has a simple zero at $h=d/2$, thus reducing the order of the pole.\footnote{Unless $\D_1+\D_2=d/2$, for which $\hat{\r}_{\D_i}(h,0)$ has a simple pole at $h=d/2$, rather than a simple zero. However, in the examples we know of (see the quartic melonic model and the fishnet model in the next section), this case corresponds to a transition occurring in the free limit of the theory, thus at vanishing eigenvalue, and hence we have again a simple pole in total.}
The restriction to a simple pole with $r_\star\neq 0$ serves to steer clear of such degenerate situation.
Hypothesis~\ref{hyp2} is actually slightly redundant on this point, as the property $k_{\s}(\htilde,J)=k_{\s}(h,J)$, following from the reality condition, implies that $h_\star=d/2$ cannot be a simple zero, but it is perhaps more clear if stated explicitly.

%%%%%%%%%%%%%%
%%%%%%%%%%%%%%

\

%%%%%%%%%%%%%%%%%%%%%%
\paragraph{Putting things together.}
%%%%%%%%%%%%%%%%%%%%%%
In order to test the stability of the $G_\star$ solution (or ``vacuum'') we need to consider the effective action at quadratic order in the fluctuations $\d G= G-G_\star$, as in \eqref{eq:Gamma-quadratic}. 
Using \eqref{eq:Hess-K}, expressing the fluctuations as in \eqref{eq:f-transform-real} with the identification $f(X_1,X_2)=\d G(X_1,X_2)$,  and then using the eigenvalue equation \eqref{eq:eigV} and the orthogonality relations \eqref{eq:V-orthonorm}, we find:
\be \label{eq:Hess-K-final}
\begin{split}
\mathbf{\G}[G] - \mathbf{F} \simeq  &\, \f14 \int_{X_1\ldots X_6} \d G (X_1,X_2)  G_\star^{-1}(X_1,X_5)G_\star^{-1}(X_2,X_6) \\
& \qquad\qquad\qquad\quad \times \left(\mathbb{I}-K\right)(X_5,X_6,X_3,X_4)  \d G (X_3,X_4) \\
= &\, \f18 \sum_{J\in \mathbb{N}_0}    \int_{\cP}
 \frac{{\rm d}h}{2\pi \im}
   \, \r(h,J)\,  \sum_{\s}\,  (1-k_{\s}(h,J)) 
    \int \dd  z  F_{\htilde;\s}^{\m_1\cdots \m_{J}}(z)  F_{h;\s}^{\m_1\cdots \m_{J}}(z)  \,.
\end{split}
\ee

The conjugation property \eqref{eq:F-conj} implies that the integrand of the $z$-integral is a positive function.
Therefore, since $\r(h,J)$ is also a positive function on the principal series, negative contributions to the integral over $h$ can only arise from the factor $(1-k_{\s}(h,J))$. And indeed, by Hypothesis \ref{hyp2}, $(1-k_{\s}(h,J))$ must change sign on the integration contour around the simple root $h_\star\in\cP$,  if the kernel eigenvalue is real on the principal series.\footnote{In Hypothesis~\ref{hyp2} we have assumed a real Bethe-Salpeter kernel, so that it is diagonalizable, with real eigenvalues. However, the instability argument should apply also to a complex Bethe-Salpeter kernel, as long as it is diagonalizable and $(1-k_{\s}(h,J))$ is not purely imaginary for $h$ in a subinterval of the principal series containing $h_\star$.}
A pictorial example of the signs structure is given in Figure~\ref{fig:merge-transition}.
Notice once more that since $k_{\s}(\htilde,J)=k_{\s}(h,J)$, we are guaranteed to find negative values of $(1-k_{\s}(h,J))$ only for $r_\star\neq 0$.
As a consequence we can find fluctuations such that the right-hand side of \eqref{eq:Hess-K-final} is negative, e.g.\ fluctuations
with a transform $F_{h;\s}^{\m_1\cdots \m_{J}}(z)$ peaked on some value of $h$ near $h_\star$ at which $(1-k_{\s}(h,J))<0$.

In summary, we have proved the following proposition, which is a more precise version of Claim~\ref{claim1}:

\begin{proposition} \label{prop1}
Given Hypothesis~\ref{hyp1} and \ref{hyp2},  there exist perturbations $\d G(X_1,X_2)\in \mathcal{V}$ such that the second variation of the 2PI effective action $\mathbf{\G}[G]$ around the solution $G_\star(X_1,X_2)$ is negative.
Therefore, the conformal solution $G_\star(X_1,X_2)$ is unstable.
\end{proposition}

Notice that from \eqref{eq:V-complet} on we have assumed $d>1$, otherwise the measure should be modified and the contributions from the discrete series representations should be included as well  \cite{Maldacena:2016hyu}. However, as the principal series remains part of the representation, the conclusion of Proposition~\ref{prop1} is unaffected: one can in fact choose fluctuations as above with the additional condition that they have a vanishing projection on the discrete series, and thus find again that the right-hand side of \eqref{eq:Hess-K-final} is negative.

We conclude this section with the following small remark. Besides the positive and negative modes, the Hessian of the effective action obviously admits a zero mode, the $V$-function with $h=h_\star$. One should not be tempted to interpret this as the dilaton, the Goldstone mode of broken conformal invariance, because the dilaton should be identified with the flat direction of a stable vacuum, the existence of which we have not proved.
In our case, the zero mode is just a consequence of having a continuous spectrum with both negative and positive eigenvalues.
For contrast, one should compare to the SYK model, where there are no negative modes, and there is one isolated zero mode on the discrete series (at $h=2$) that can then be interpreted as the Goldstone mode for the spontaneous breaking of the full conformal symmetry (time reparametrization invariance) down to $SL(2,\mathbb{R})$  \cite{Maldacena:2016hyu}.  We should also point out that while in the SYK model such zero mode in the discrete series representation leads to a problem when inverting the Hessian to write the four-point function, and one is forced to move away from the conformal IR limit, this is not the case for the zero mode in the principal series, as the integral in \eqref{eq:CPW-rep} will in general be defined by a slight contour deformation, as we explained above.

%\newpage
%%%%%%%%%%%%%%%%%%%%%%%%%%%%%
\section{Examples}
\label{sec:examples}
%%%%%%%%%%%%%%%%%%%%%%%%%%%%%

As mentioned in the introduction, operators with scaling dimension of the type $h=h_\star\equiv \f{d}{2}+\im  r_\star$  have been found in a variety of models.
In this Section, we consider two recent classes of large-$N$ examples, which allow a direct evaluation of the eigenvalues of the Bethe-Salpeter kernel: theories dominated by melonic diagrams \cite{Murugan:2017eto,Giombi:2017dtl,Prakash:2017hwq,Giombi:2018qgp,Kim:2019upg,Benedetti:2019eyl,Benedetti:2019rja,Klebanov:2020kck,Benedetti:2020iku}, and theories dominated by fishnet diagrams \cite{Grabner:2017pgm,Kazakov:2018qbr,Gromov:2018hut}.
While these two classes of models are fundamentally different because of the different type of diagrams that dominate the large-$N$ limit, it turns out that at the level of the four-point function the fishnet models are described by the same diagrams as melonic models with quartic interactions: the diagrams on the right-hand side of Figure~\ref{fig:F-ladder} are in both cases true ladders, with double-edge rungs.
Their two-point function diagrams are however different, which results in a different behavior at $d=4$; for $d<4$, one can instead consider long-range versions for both models (i.e.\ a with a fractional power of the Laplacian in the kinetic term) and obtain similar renormalization group equations for the double-trace interactions.
We here consider only long-range models, as they have the advantage that the SD equations do not require an IR limit, and that there is an exactly marginal coupling that we can tune at will. From the point of view of the instability, the behavior of the short-range models (fishnet model in $d=4$, or melonic models in an $\eps$ expansion near the critical dimension) is qualitatively similar, and when the marginal coupling is absent its role is essentially replaced by the $\eps$ parameter.

The melonic models, unlike the fishnet models, have been constructed also for interactions of a different order than four, e.g.\ sextic \cite{Giombi:2018qgp,Benedetti:2019rja} or cubic interactions \cite{Benedetti:2020iku}. 
We will start with the latter, as in that case the conformal dimension of the fundamental fields is $d/3$, which thus satisfies the condition \eqref{eq:Deltas-conditions}, while models with higher interactions require an analytic continuation. 
We will then discuss quartic models, for which the transition to instability happens at the free theory, as the exactly marginal coupling goes from purely imaginary to real; as a consequence one could try to look for a stable solution in the small-coupling limit, something that we hope to address in the near future.
Lastly, we will consider two-flavor (tensor and fishnet) models in which the instability is associated to a spontaneous symmetry breaking.

%%%%%%%%%%%%%%%%%%%%%%
\subsection{Amit-Roginsky model}
%%%%%%%%%%%%%%%%%%%%%%

The Amit-Roginsky (AR) model, introduced in \cite{Amit:1979ev} and recently revisited in \cite{Benedetti:2020iku}, is a bosonic model of $N$ complex scalar fields $\phi_m$, with $m=1\ldots N$, in an irreducible representation of $SO(3)$ of dimension $N=2j+1$, with even $j$, and with a cubic interaction mediated by a Wigner $3jm$ symbol. 
We consider a variation of the model with real scalars and with a  long-range quadratic part of the action.
The action reads:
\be \label{eq:AR}
\begin{split}
S[\phi] = \int \dd x \; &\left(\f12 \sum_m \left(\phi^m (-\p^2)^\z\phi_m + \l_2\, \phi_m \phi^m\right) \right.\\
&\left.+\sum_{m_1,m_2,m_3} \f{\l}{3!}\sqrt{2j+1}\tj{j}{j}{j}{m_1}{m_2}{m_3} \phi^{m_1}\phi^{m_2}\phi^{m_3} \right) \,,
\end{split}
\ee
where $\tj{j}{j}{j}{m_1}{m_2}{m_3}$ is the $3jm$ symbol, and the indices are raised and lowered by the $SO(3)$ invariant metric (see the appendix of \cite{Benedetti:2019sop} for a brief compendium of useful formulas written in the same notation used here). We choose $\z=d/6$, so that the interaction is marginal, and $d<6$ so that the kinetic term is non-local, the propagator is unitary, and local derivative interactions are irrelevant.
The noninteger power of the Laplacian should be understood as the integral kernel
\be \label{eq:free-C}
C^{-1}(x,y) = (-\p^2)^{\z} \equiv  \f{2^{2\z}\G\left(\f{d+2\z}{2}\right)}{\pi^{d/2}|\G(-\z)|} \f{1}{|x-y|^{d+2\z}} \,.
\ee

Because of the global $SO(3)$ symmetry, and because when looking for a conformal solution we must assume that the symmetry is not broken, the index structure is constrained. Therefore, in building the 2PI effective action we restrict to invariant solutions and fluctuations, i.e.\ we take
$G^m_{m'}(x,y) = \d^m_{m'} G(x,y)$, by which we thus define the index-free part $G(x,y)$.

\begin{figure}[htbp]
\centering
\includegraphics[width=0.2\textwidth]{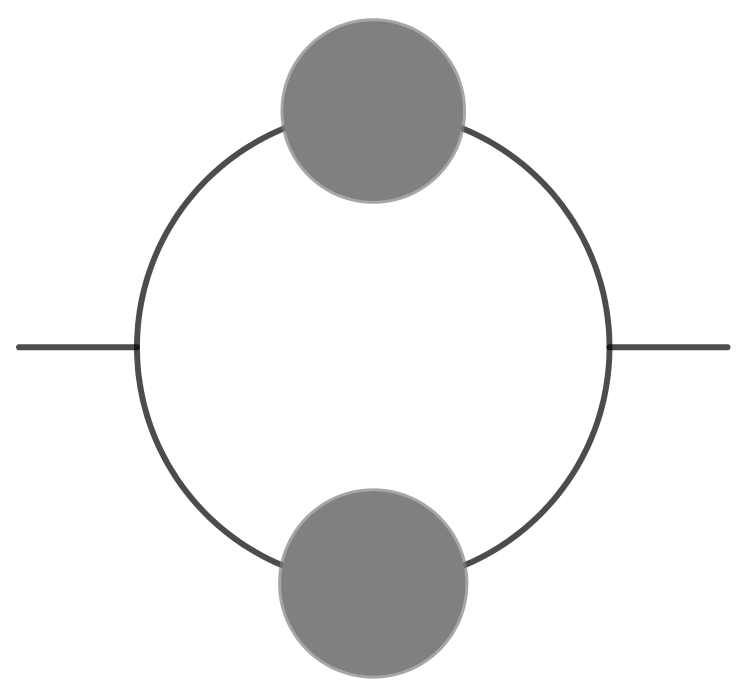}\\
\caption{The self-energy of the Amit-Roginsky model. Full propagators are represented by the gray blobs.}
\label{fig:self-energy}
\end{figure}

In the large-$N$ limit, the AR model is dominated by melonic Feynman diagrams \cite{Amit:1979ev,Benedetti:2020iku},\footnote{A proper proof is missing, but the claim is supported by numerical tests.} and the same must be true for the diagrams defining the 2PI effective action.
The melonic dominance, combined with the 2PI restriction, leads to the effective action
\be \label{eq:AR-Gamma_2PI}
\mathbf{\G}[G] = N\left(\f12 \Tr[(-\p^2)^\z G]  + \f12 \Tr[\ln G^{-1}] + \f{\l_2}{2} \int_x G(x,x) -\f{\l^2}{12}\int_{x,y} G(x,y)^3\right)\,.
\ee
From this we obtain the Schwinger-Dyson equation, which in momentum space reads
\be \label{eq:AR-SDeq}
G(p)^{-1} =  p^{d/3} + \l_2  - \f{\l^2}{2} \int_q G(q) G(p+q) \,,
\ee
where $\Sigma(p^2)=\f{\l^2}{2} \int_q G(q) G(p+q)$ is the melonic self-energy (see Figure~\ref{fig:self-energy}).
%%%%%%%%%%%%%%%%
%%%%%%%%%%%%%%%%
The mass coupling includes a counterterm canceling the $p$-independent divergent part of the on-shell self-energy, namely $\l_2 = \l_2^c + g_2$, with $\l_2^c = \f{\l^2}{2} \int_q G_\star(q)^2$.  No wave function renormalization is needed, because of the non-local nature of the propagator.
Tuning the renormalized mass to the critical value $g_2 = 0$, one finds the solution $G_\star(p)= \cZ p^{-d/3}$, valid at all scales.
The proportionality coefficient $\cZ$, written explicitly in \cite{Benedetti:2020iku}, solves a cubic equation, and it resums exactly all the melonic insertions.
Therefore, we have an explicit realization of the SD equation \eqref{eq:SDeq-2pt}, with an exact conformal solution, 
\be \label{eq:AR-Gsol}
G_\star(x,y) = \cZ   \int \f{\dd p}{(2\pi)^d}\f{e^{ip(x-y)}}{p^{2\zeta} }= \cZ  \f{\G\left(\Delta_{\phi}\right)}{2^{d-2\D_{\phi}}\pi^{d/2}\G(\f{d}{2}-\D_{\phi})} \f{1}{|x-y|^{2\D_{\phi}}}\,,
\ee
where the $N$ fundamental fields have all the same dimension $\D_\phi = d/2-\z = d/3$.
The normalization in \eqref{eq:G_star} is achieved by switching to the rescaled fields $\tilde{\phi}_m = \phi_m\, 2^{\z/2}/\cZ^{1/2}$.

In the large-$N$ limit the cubic interaction is not renormalized, and since the kinetic term is non-local there is also no wave-function renormalization; therefore, the beta function of $\l$ is identically zero, i.e.\ $\l$ is an exactly marginal coupling.

In the short-range version of the model, the free part of \eqref{eq:AR-SDeq} has a standard power $p^2$, rather than $p^{d/3}$, hence a solution with the same scaling as in the long-range model is found only by taking the limit $p\to 0$, and by tuning the coupling (which is not marginal in this case) to its fixed point.
This is the kind of limits we alluded to in footnote \ref{foot:limitSD}. The advantage of the long-range model is that they are not required.

By $SO(3)$ representation theory (see Chapter III, section 13 of \cite{Yutsis:1962vcy}), the full Bethe-Salpeter kernel can be decomposed as
\be
\begin{split}
K_{m_1 m_2, m_3 m_4} (x_1,x_2,x_3,x_4) = \sum_{\ell=0}^{2j} \sum_{m=-\ell}^{\ell} \, (-1)^{\ell-m} &(2\ell+1)  K_{\ell,m} (x_1,x_2,x_3,x_4) \\
&\times \tj{j}{j}{\ell}{m_1}{m_2}{m} \tj{j}{j}{\ell}{m_3}{m_4}{-m}\,,
\end{split}
\ee
and therefore we identify the matrix structure of the $V$-functions in \eqref{eq:V-def}, such that they form an eigenbasis of $K$, to be:
\be
E^{\s,J}_{m_1 m_2} = \sqrt{2\ell+1} \tj{j}{j}{\ell}{m_1}{m_2}{m}  \,,
\ee
with $\s=(\ell,m)$, and $\ell$ restricted to be even (odd) if $J$ is even (odd).

Having restricted also the fluctuations in the 2PI effective action to be diagonal, we only need the trace of the Bethe-Salpeter kernel.
Tracing the pairs $(1,2)$ and $(3,4)$, projects $K$ on the $\ell=m=0$ component:
\be
K_{m}{}^{ m}{}_{,m'}{}^{ m'} (x_1,x_2,x_3,x_4) = N \, K_{0,0} (x_1,x_2,x_3,x_4) \,.
\ee

\begin{figure}[htbp]
\centering
\includegraphics[width=.6\textwidth]{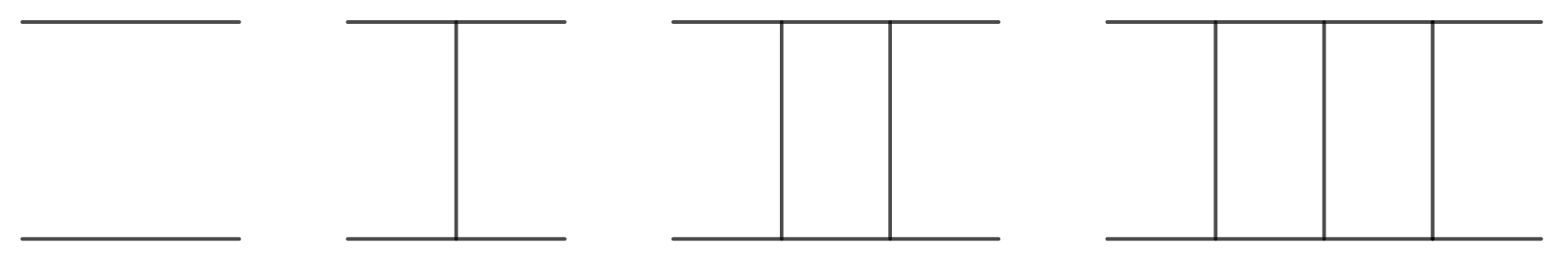}\\
\caption{The first four contributions to the ladder expansion of $\cF_s$ in the Amit-Roginsky model, where again we omit the crossed diagrams with $x_3$ and $x_4$ exchanged.}
\label{fig:AR-ladder}
\end{figure}

In the melonic limit, we have
\be \label{eq:AR-K00}
K_{0,0}(x_1,x_2,x_3,x_4) = G_\star(x_1,x_3) G_\star(x_2,x_4) G_\star(x_3,x_4) \,,
\ee
which generates the ladder expansion of $\cF_s$, depicted in Figure~\ref{fig:AR-ladder}.
%%%%%%%%%%%%%%%%
%%%%%%%%%%%%%%%%
The conformal invariant two-point function \eqref{eq:AR-Gsol} and four-point function \eqref{eq:AR-K00}, obtained from the 2PI effective action, provide an example of the situation imagined in Hypothesis~\ref{hyp1}.

The eigenvalues of the kernel $K_{0,0}$ are:
\be  \label{eq:AR-k}
k_{0,0}(h,J)=2 g^2 \f{1}{(4\pi)^{d/2}} 
\f{\G(\f{d}{3}) \G\left(\f{d}{3}-\f{h-J}{2}\right) \G\left(\f{h+J}{2}-\f{d}{6}\right) }{ \G(\f{d}{6}) \G\left(\f{2d}{3}-\f{h-J}{2}\right) \G\left(\f{h+J}{2}+\f{d}{6}\right)}\,.
\ee 
where for convenience we have introduced the rescaled coupling $g=\l \cZ^{3/2}$.
Notice that they are meromorphic functions of $h$ in the whole complex plane, and they are real for $h\in \cP$, hence the first half of Hypothesis~\ref{hyp2} is satisfied.

\begin{figure}[htbp]
 \centering
        \begin{minipage}{0.45\textwidth}
            \centering 
             \includegraphics[width=0.98\textwidth]{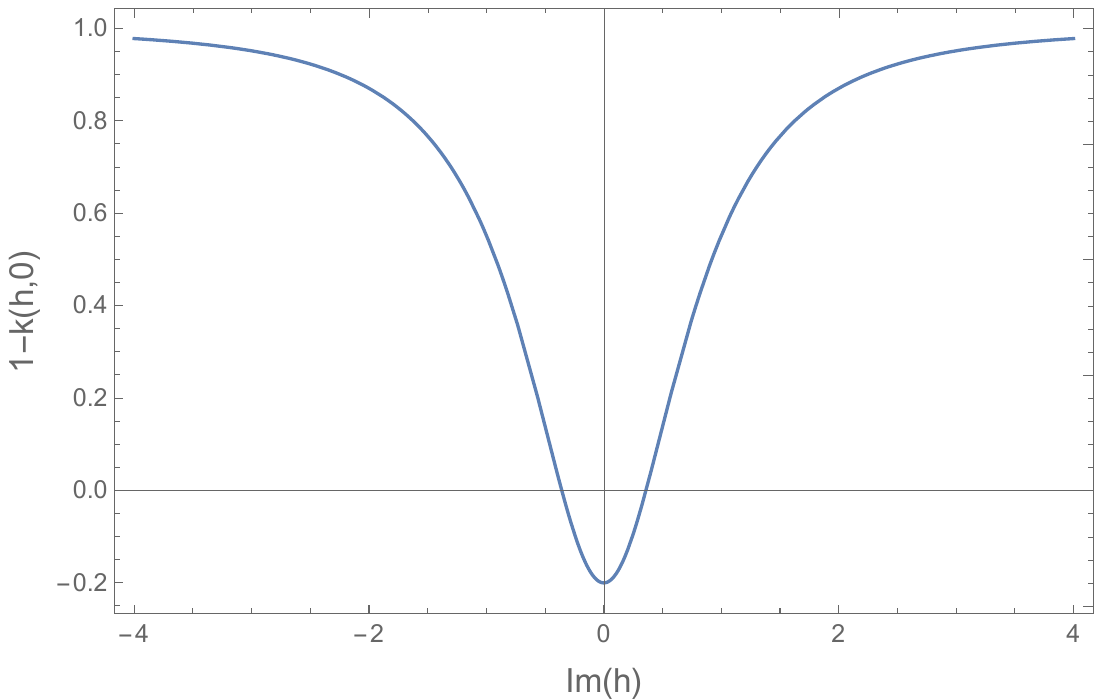}
        \end{minipage}
        \hspace{0.01\textwidth}
        \begin{minipage}{0.45\textwidth}
            \centering
            \includegraphics[width=0.999\textwidth]{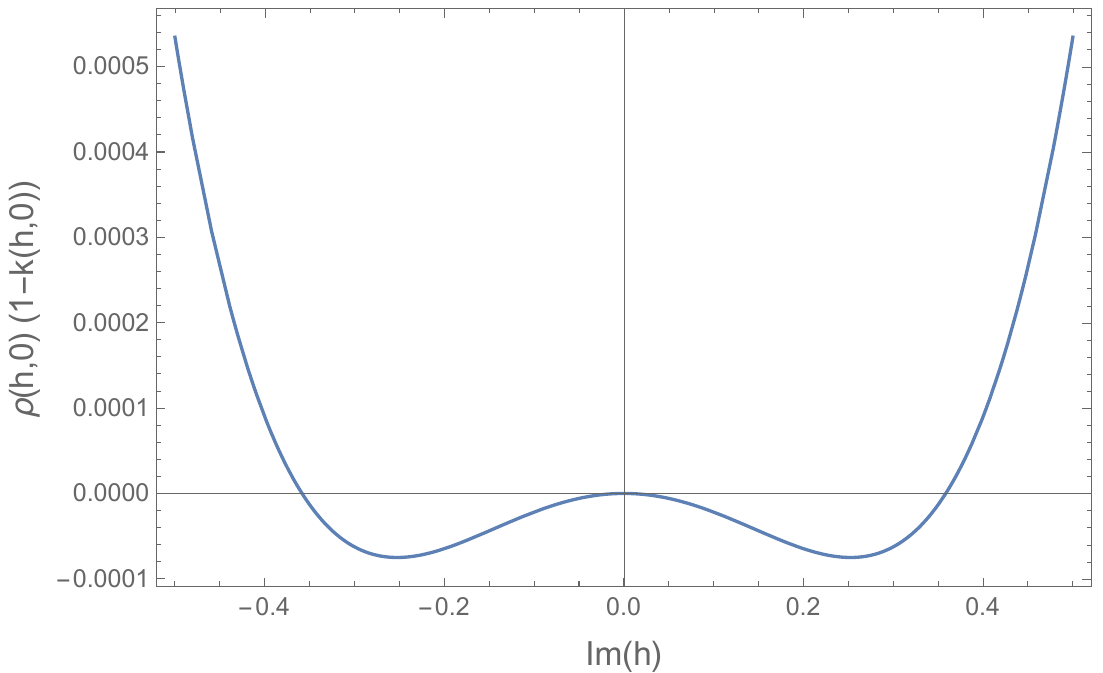}
        \end{minipage}
\caption{Plot of $1-k_{0,0}(h,0)$ (left) and $\r(h,0)(1-k_{0,0}(h,0))$ (right) at $d=5$ and $g=10$, along the principal series \eqref{eq:principal-series}, for the Amit-Roginsky model, whose kernel eigenvalue is given in \eqref{eq:AR-k}.}
\label{fig:AR-plot}
\end{figure}

By analytically solving the equation $k_{0,0}(h,J)=1$  at small coupling, one finds a set of non-degenerate solutions $h(n,J)=2d/3+2n+J+z_{n,J}$, for $n\in\mathbb{N}_0$, with anomalous dimension $z_n$ of order $g^2$.
Numerically, one can go also to large values of $g$, and 
it was found in \cite{Benedetti:2020iku} that for any $d<6$ there is a merging of $h(0,0)$ with its shadow $\tilde{h}(0,0) = d-h(0,0)$ at a critical value of $g$, and beyond that the two solutions move along the principal series.
Such merging occurs at $g=  9.17$  for  $d=5$, and Figure~\ref{fig:AR-plot} shows how at larger values of $g$ it implies the existence of a region with $1-k_{0,0}(h,0)<0$ along the principal series, thus providing a clear example of Proposition~\ref{prop1}.
In other dimensions a similar situation is found, with the critical value of $g$ decreasing together with $d$.
%%%%%%%%%%%%%%%%
%%%%%%%%%%%%%%%%

%%%%%%%%%%%%%%%%%%%%%%
\subsection{$O(N)^3$ tensor model with quartic or higher interactions}
%%%%%%%%%%%%%%%%%%%%%%

Tensor models with $O(N)^3$ symmetry and standard short-range kinetic term have been studied in one or more dimensions in \cite{Giombi:2017dtl,Prakash:2017hwq,Giombi:2018qgp,Kim:2019upg,Benedetti:2019rja,Klebanov:2016xxf,Bulycheva:2017ilt,Gubser:2017qed,Chang:2018sve,Popov:2019nja,Lettera:2020uay} (see also \cite{Klebanov:2018fzb} for a review).
Bosonic models can be written with $q$-valent interactions, in principle for any even $q>2$ but in practice typically for $q=4$ or $6$. They can be studied as usual in an $\epsilon$ expansion near the upper critical dimension, that is, at $d=d_c-\eps=2q/(q-2)-\eps$.
At large-$N$, the melonic dominance gives rise to a non-trivial fixed point, at which the tensor fields have scaling dimension $d/q$, which for $q> 4$ violates  \eqref{eq:Deltas-conditions}. Therefore, models with higher-order interactions, such as the sextic ($q=6$) models studied in \cite{Giombi:2018qgp,Benedetti:2019rja}, need some special care with their analytic continuation (see the discussion  in \cite{Aharony:2020omh} or in the appendix of \cite{Benedetti:2020iku}).
One general feature is the appearance at $\eps>\eps_c>0$ (for $q=6$) or at all $\eps>0$ (for $q=4$) of a complex scaling dimension with value in the principal series, and it can be checked that the argument behind Proposition~\ref{prop1} applies.
However, short-range models are not the most convenient examples for the following reasons: first, the SD equations in the short-range models are only solvable in the IR limit; second, in the $q=4$ short-range model there is no stable conformal phase for $\eps>0$, and the stable one at $\eps<0$ describes a non-unitary UV theory ($d/4<d/2-1$ for $d>4$, i.e.\ the unitarity bound is violated).

It is more convenient to illustrate the instability in long-range models.
The long-range quartic $O(N)^3$ model has been introduce and studied in \cite{Benedetti:2019eyl,Benedetti:2019ikb,Benedetti:2020yvb,Benedetti:2020sye}  (see also \cite{Benedetti:2020seh} for a review).

The fundamental field is a real rank-$3$ tensor field, $\phi_{a^1a^2 a^3}(x)$,
transforming under $O(N)^3$ with indices distinguished by the position (typically labelled by a color).  Denoting $\mba = (a^1,a^2,a^3)$, the action of the model is:
\be	\label{eq:ON3-action}
\begin{split}
S[\phi] &=  \frac{1}{2} \int \dd x \, \phi_{\mba} (  - \partial^2)^{\zeta}\phi_{\mba} + 
	\frac{ m^{2\zeta}}{2} \int \dd x \, \phi_{\mba}  \phi_{\mba} \\
	&\qquad + \frac{1}{4} \int \dd x \, \left[ \lambda \hat{\delta}^t_{\mba \mbb\mbc\mbd} + \lambda_1 \hat{P}^{(1)}_{\mba\mbb; \mbc\mbd}
	+ \lambda_2 \hat{P}^{(2)}_{\mba\mbb; \mbc\mbd } \right] \phi_{\mba} \phi_{\mbb} \phi_{\mbc} \phi_{\mbd} \, , 
\end{split}
\ee
where repeated tensor indices are summed over $a^i = 1, \cdots, N$ and we introduced the projectors:
	\begin{equation}
		\hat{P}^{(1)}_{\mba\mbb; \mbc\mbd} \, = \, 3 (\hat{\delta}^p_{\mba\mbb;\mbc\mbd} - \hat{\delta}^d_{\mba\mbb;\mbc\mbd}) \, , \qquad 
		\hat{P}^{(2)}_{\mba\mbb; \mbc\mbd} \, = \, \hat{\delta}^d_{\mba\mbb;\mbc\mbd} \, .
	\end{equation}
and the rescaled operators:
\be \label{eq:deltas}
\hat{\delta}^t_{\mba\mbb\mbc\mbd}=\frac{1}{N^{3/2}} \, \delta^t_{\mba\mbb\mbc\mbd} \,,
\quad \hat{\delta}^p_{\mba\mbb;\mbc\mbd}=\frac{1}{N^{2}} \, \delta^p_{\mba\mbb;\mbc\mbd}\,,
\quad \hat{\delta}^d_{\mba\mbb;\mbc\mbd}=\frac{1}{N^{3}} \, \delta^d_{\mba\mbb;\mbc\mbd}\, ,
\ee
with
\be
\begin{split} \label{eq:deltas-nohat}
&\delta^t_{\mba \mbb\mbc\mbd}  = \delta_{a^1 b^1}  \delta_{c^1 d^1} \delta_{a^2 c^2}  \delta_{b^2 d^2 } \delta_{a^3 d^3}   \delta_{b^3 c^3} \, , \\
	\delta^p_{\mba\mbb; \mbc\mbd } &= \frac{1}{3} \sum_{i=1}^3  \delta_{a^ic^i} \delta_{b^id^i} \prod_{j\neq i}  \delta_{a^jb^j}  \delta_{c^jd^j} \,,
	\qquad  \delta^d_{\mba\mbb; \mbc\mbd }  = \delta_{\mba \mbb}  \delta_{\mbc \mbd} \,.
\end{split}
\ee
Here $t$ stands for \emph{tetrahedron},  $p$ for \emph{pillow}, and $d$ for \emph{double-trace}. Such names refer to the graphical representation of the respective pattern of contraction of indices.
The faithfully acting symmetry group of the model is $S_3 \times O(N)^3/\mathbb{Z}_2^2$, where the quotient by $\mathbb{Z}_2^2$ is to eliminate the redundancy of having a $\mathbb{Z}_2$ as subgroup in each $O(N)$, and $S_3$ is the permutation group acting on the indices. Symmetry under the latter is often called ``color symmetry", due to using color as a label distinguishing the indices, and not to be confused with the gauge theory terminology from quantum chromodynamics.
The power of the Laplacian is taken to be  $\z=d/4$, so that  $\D_{\phi}=d/4$ and the quartic couplings are marginal by power counting. Moreover, one assumes $d<4$ so that the kinetic term is non-local, the propagator is unitary, and local derivative interactions are irrelevant.

We can again simplify our task by building the 2PI effective action directly in the symmetric phase, 
\be \label{eq:diag-ansatz}
G_{\mba\mbb}(x,y) = G(x,y)\, \d_{\mba\mbb}\,.
\ee
The indices then just lead to the standard counting of the powers of $N$ of $O(N)^3$ tensor models \cite{Carrozza:2015adg,Klebanov:2016xxf}. For large-$N$, the leading order effective action is of order $N^3$, and reads
\be \label{eq:ON3-Gamma_2PI}
\begin{split}
\mathbf{\G}[G] = & N^3\left(\f12 \Tr[(-\p^2)^\z G]  + \f12 \Tr[\ln G^{-1}] + \f{m^{2\zeta}}{2} \int_x G(x,x) \right.\\
&\qquad\quad \left.+ \f{\l_2}{4} \int_x G(x,x)^2 -\f{\l^2}{8}\int_{x,y} G(x,y)^4\right)\,.
\end{split}
\ee
The SD equations are then similar to \eqref{eq:AR-SDeq}, but with an extra edge in the self-energy, plus some tadpole terms, and of course a different value of $\z$:
\be \label{eq:ON3-SDeq}
G(p)^{-1} =  p^{d/2} + m^{d/2} +\l_2 \int_{q}    G(q) - \lambda^2    \int_{q_1,q_2}    G(q_1)  G(q_2)  G( p +q_1 + q_2  ) \,.
\ee
Tuning the bare mass to cancel the tadpole and the $p=0$ contribution of the melonic integral, an exact conformal solution is found, $G_\star(p)=\cZ p^{-d/2}$, where $\cZ$ now solves a quartic equation, again taking into account all the melonic two-point insertions.

In the large-$N$ limit the tetrahedron interaction is not renormalized, and since the kinetic term is non-local there is also no wave-function renormalization; therefore, the tetrahedron beta function is identically zero, i.e.\ $g$ is an exactly marginal coupling. However, unlike in the Amit-Roginsky model, where there was a unique cubic interaction, here there are two other couplings to consider. It turns out that the beta functions of $g_1$ and $g_2$ (the renormalized $\l_1$ and $\l_2$) are decoupled from each other, and they are quadratic polynomials in  $g_1$ and $g_2$, respectively, with coefficients that depend parametrically on $g$. 
For real $g$ the discriminants of their fixed point equations are negative, hence the fixed points are complex, while for purely imaginary $g$ the discriminants are positive, hence the fixed points are real.
This is a concrete realization of the general scenario of running of double-trace couplings discussed in \cite{Pomoni:2008de}, although the interaction given by the $\hat{P}^{(1)}$ projector (essentially the  pillow interaction) is strictly speaking not a double-trace interaction.

\begin{figure}[ht]
\begin{center}
$$K = 3\l^2\, \vcenter{\hbox{\includegraphics[width=1.8cm]{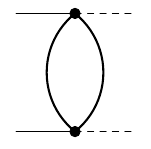}}} - \l_2 \, \vcenter{\hbox{\includegraphics[width=1.8cm]{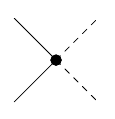}}}$$
 \caption{Graphical representation of the kernel \eqref{eq:ON3-K2} of the quartic $O(N)^3$ model. Solid lines represent full two-point functions, while dashed lines represent amputated external legs.} \label{fig:kernel}
 \end{center}
\end{figure}

The second variation of \eqref{eq:ON3-Gamma_2PI}, divided by $N^3$, leads to the Bethe-Salpeter kernel\footnote{More generally, the invariance of the model implies that the index structure of the Bethe-Salpeter kernel is built just from Kronecker delta functions, and that it can be decomposed into projectors. The $O(N)^3$ invariance alone implies that there are in total 27 possible structures (from the three possible pairwise identifications of the four indices of each color), while symmetrization on $(1,2)$ and the color symmetry reduce them to six (essentially the tetrahedron, three orientations of the pillow, and two of the double trace). Orthogonal projectors can then be constructed by hand, of which only two are needed in the large-$N$ limit, namely $\hat{P}^{(1)}$ and $\hat{P}^{(2)}$ \cite{Benedetti:2019eyl}. We will not need any of this here, as the symmetric restriction on the two-point function and its fluctuations implies that we only keep the trace-projector part of the kernel.}
\be \label{eq:ON3-K2}
K(x_1,x_2,x_3,x_4) = G_\star(x_1,x_3) G_\star(x_2,x_4) \left( 3\l^2 G_\star(x_3,x_4)^2 - \l_2 \d(x_3-x_4)\right)  \,,
\ee
which is represented in Figure~\ref{fig:kernel}. Its geometric series (i.e.\ replacing Figure~\ref{fig:kernel} in the gray rectangles of Figure~\ref{fig:F-ladder}) gives rise to a sum of mixed chain-ladder diagrams, where the ladders have double-edge rungs.
%%%%%%%%%%%%%
%%%%%%%%%%%%%

The eigenvalues of the Bethe-Salpeter kernel are
\begin{equation}
k(h,J)=\f{3g^2}{(4\pi)^d}
 \frac{\Gamma(-\frac{d}{4}+\frac{h+J}{2})\Gamma(\frac{d}{4}-\frac{h-J}{2})}{\Gamma(\frac{3d}{4}-\frac{h-J}{2})\Gamma(\frac{d}{4}+\frac{h+J}{2})} \,,
\label{eq:ON3-k}
\end{equation}
where we have introduced the rescaled coupling $g=\l \cZ^{2}$. The relation between $\l$ and $g$, for real $g$ is invertible up to the critical value
\be \label{eq:ON3-gc}
g_c= (4\pi)^{d/2}\left( \f{ \f{d}{4}\, \G(3\f{d}{4}) }{ \G(1-\f{d}{4}) } \right)^{1/2}\,.
\ee
For $\D_{\phi}=d/4$, because of the $\d(x_3-x_4)$ in \eqref{eq:ON3-K2},  when evaluating the integral in the eigenvalue equation \eqref{eq:eigV}  the  $\l_2$ term gives a vanishing contribution for ${\rm Re}(h)>d/2$, where we expect to find the physical dimensions, and a divergent one for ${\rm Re}(h)<d/2$; for $h\in \cP$, which is the important case corresponding to the basis of eigenfunctions, the result is indeterminate. This can be regulated by the same analytic regularization used for the perturbative computations in \cite{Benedetti:2020yvb,Benedetti:2020sye}, that is, taking $\z=(d+\eps)/4$, or equivalently $\D_{\phi}=(d-\eps)/4$; then, the $\l_2$ term does not contribute to the kernel eigenvalue, consistently with the ${\rm Re}(h)>d/2$ half-plane where we wish to move the contour in \eqref{eq:CPW-rep}.

One finds two types of solutions of the equation $k(h,J)=1$ at small renormalized tetrahedral coupling $g$. The first type,
\begin{equation} \label{eq:h_pm}
h_{\pm} = \frac{d}{2} \pm 2 \frac{  \Gamma(d/4)^2  }{ \Gamma(d/2) } \sqrt{ - 3g^2  } \left(1 + { O}(g^2)\right) \, ,
\end{equation}
only exists for the $J=0$ case, and can be understood as the dimension of the bilinear operator $\cO_{0,0}\sim\phi_{abc}\phi_{abc}$. 
The second type of solution is
\begin{equation} \label{eq:h_mJ}
 h_{n,J} = \frac{d}{2} + J + 2n +  \frac{ \Gamma(d/4)^4 
     \Gamma(n + J)  \Gamma(n+1 - \frac{d}{2}) \sin\left(\frac{\pi d}{2} \right)  }{ \Gamma(\frac{d}{2} + J + n ) \Gamma(n+1) \; \pi } 6g^2  + 
     { O}(g^4) \, ,
\end{equation}
with $n,J \in \mathbb{N}_0$, but not simultaneously zero. 
In the free limit $g=0$, we recover the classical dimensions $\frac{d}{2}+J+2n$ of the primary bilinear operators with arbitrary spin $J$, schematically of the form
\be
\cO_{n,J} \sim  \phi_{abc} \partial_{\mu_1}\dots \partial_{\mu_J} (\partial^2)^{n}\phi_{abc}  \,.
\ee

While the second type of solution is real (at all orders in $g$) for both real and purely imaginary tetrahedral coupling,
the first type is real only for purely imaginary $g$; for real $g$ it belongs to the principal series. 
Therefore, we have an interesting situation in which for real $g$ we find that $h_{\pm}\in \cP$ already for infinitesimal coupling, so that the instability and perhaps the existence of a different (stable) vacuum could be studied perturbatively.
The two sign choices are related by a shadow transform, $h_-=d-h_+$, and their merging happens in the free theory limit, i.e.\ at $g=0$.
For imaginary $g$, it can be checked that the physical solution at the IR fixed point is $h_+$, while $h_-$ is the physical dimension at the UV fixed point  \cite{Benedetti:2019eyl}. The flow between the two is driven by the double-trace perturbation $\cO_{0,0}^2$, as by the general argument of \cite{Gubser:2002vv}. For real $g$, the two sign choices still characterize the physical dimensions at different fixed points, but since the critical exponents are purely imaginary, the two fixed points are not connected by a renormalization group trajectory.

\begin{figure}[htbp]
 \centering
        \begin{minipage}{0.45\textwidth}
            \centering 
             \includegraphics[width=0.98\textwidth]{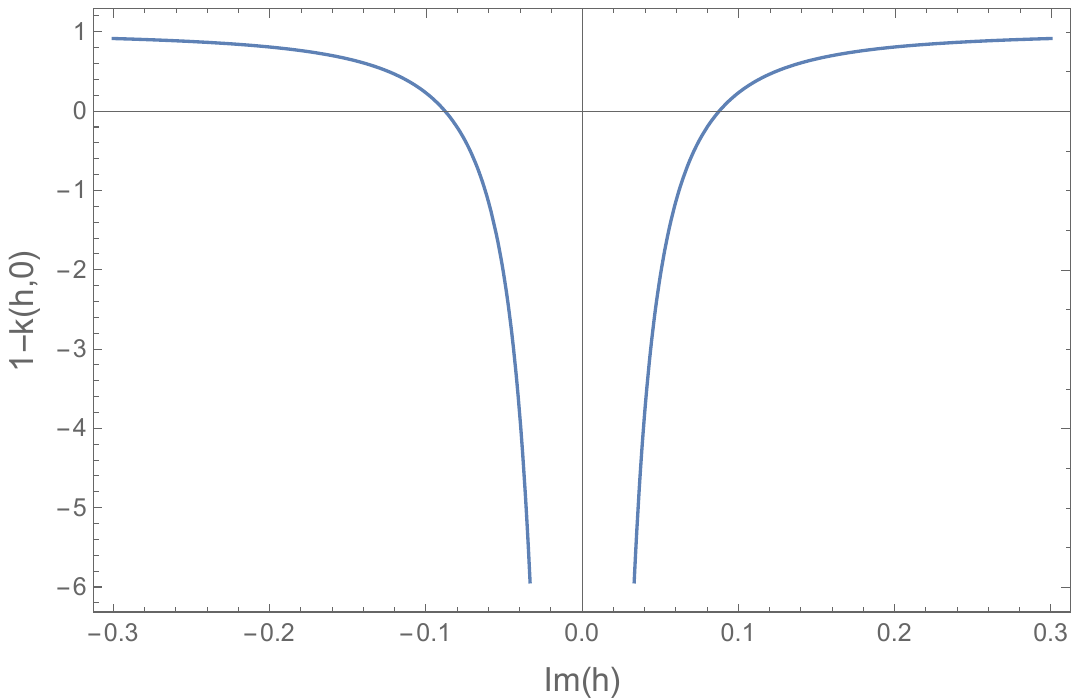}
        \end{minipage}
        \hspace{0.01\textwidth}
        \begin{minipage}{0.45\textwidth}
            \centering
            \includegraphics[width=0.995\textwidth]{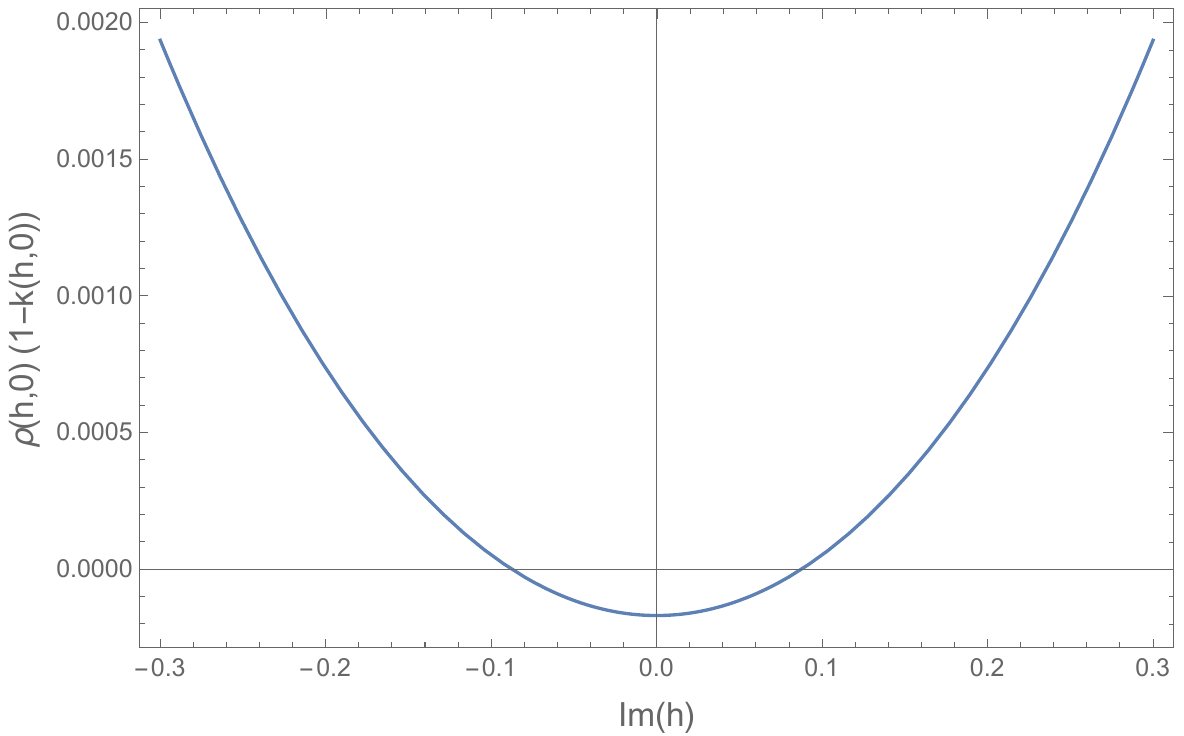}
        \end{minipage}
\caption{Plot of $1-k(h,0)$ (left) and $\r(h,0)(1-k(h,0))$ (right) at $d=3$ and $g=1<g_c=21.566$, along the principal series \eqref{eq:principal-series}, for the long-range quartic $O(N)^3$ model, whose kernel eigenvalue is given in \eqref{eq:ON3-k}.}
\label{fig:ON3-plot}
\end{figure}

Figure~\ref{fig:ON3-plot} shows how at $d=3$ for real values of $g$ the complex solution implies the existence of a region with $1-k(h,0)<0$ along the principal series, thus providing another example of Proposition~\ref{prop1}.
In other dimensions a similar situation is found.

%%%%%%%%%%%%%%%%
%%%%%%%%%%%%%%%%

Tensor models can be written for various ranks, symmetries, and order of interaction. Not all tensor models lead to melonic dominance at large-$N$: for example, in a quartic $U(N)^3$ model the tetrahedron is not allowed by the symmetry and hence the model is dominated by cactus diagrams. The question of classifying tensor models admitting a melonic limit has been addressed for example in \cite{Ferrari:2017jgw,Gubser:2018yec,Prakash:2019zia}.
For models that do have a melonic limit, and having a  $q$-valent interaction, the ladder kernel is a generalization of that on the left of Figure~\ref{fig:kernel} with the vertical double line being replaced by $q-2$ lines. Simple power counting dictates the scaling of the propagator, and for long-range models, the choice $\z = \f{d(q-2)}{2q}$, with dimension below the critical value, $d<d_c=2q/(q-2)$. The kernel eigenvalue then reads
\be  \label{eq:k_q}
k_{q}(h,J)=(q-1) g^2 \f{1}{(4\pi)^{d(q-2)/2}}\left(\f{\G(\f{d}{q})}{\G(\f{d}{2}-\f{d}{q})}\right)^{q-2}
\f{ \G\left(\f{d}{q}-\f{h-J}{2}\right) \G\left(\f{h+J}{2}-\f{d(q-2)}{2q}\right) }{ \G\left(\f{d(q-1)}{q}-\f{h-J}{2}\right) \G\left(\f{d(q-2)}{2q}+\f{h+J}{2}\right)}\,.
\ee
Although tensor models can only be constructed for even $q\geq 4$, we can formally study the above expression by analytical continuation to odd values of $q$, that could be obtained from bosonic SYK-like models or perhaps some generalization of the Amit-Roginsky model, or even to non-integer values.
We find that the equation $k(h,0)=1$ always has solutions in the principal series $\cP$ for $d<d_c$ and $g^2>g_{\rm merge}(q,d)^2\geq 0$, where $g_{\rm merge}(q,d)$ is the critical real value of the coupling at which two solutions merge at $h=d/2$. We find that $g_{\rm merge}(q,d)=0$ only for $q=4$, and asymptotically for $q\to \infty$.
The largest integer dimension below $d_c$ that is available at all values of $q$ is $d=2$, for which we find that the maximum of the curve $g_{\rm merge}(q,2)$ is reached around $q=11$.
We conclude that for any $q>2$, $0<d< d_c$, and $g^2>g_{\rm merge}(q,d)^2\geq 0$, the kernel eigenvalues \eqref{eq:k_q} satisfy Hypothesis~\ref{hyp2}, and thus Proposition~\ref{prop1} applies. 

Notice also that the kernel eigenvalue \eqref{eq:k_q} has a divergence along the principal series (precisely at $h=d/2$) only for $q=4$, which however is compensated by the double zero of the Plancherel weight, as seen in Figure~\ref{fig:ON3-plot}.
Therefore, for $q=4$ the most destabilizing perturbations are those peaked around $h=d/2$, while for $q\neq 4$ they are peaked at ${\rm Im}(h)\neq 0$, around the minima of curves like that on the right-hand panel of Figure~\ref{fig:AR-plot}.

%%%%%%%%%%%%%%%%%%%%%%
\subsection{Two-flavor fermionic tensor models in one dimension}
%%%%%%%%%%%%%%%%%%%%%%

Kim et al.\ \cite{Kim:2019upg} (see also \cite{Kim:2018dtq} for more details) have introduced a model in $d=1$ with $2N^3$ Majorana fermions $\psi_i^{abc}$, where $a,b,c=1,\ldots,N$ are $O(N)^3$ indices, and $i=1,2$ is a flavor index. 
The action is
\be \label{eq:S_Kim}
\begin{split}
S[\psi] = & \int {\rm d}\t  \sum_{i=1,2}\left( \f12 \psi_i^{\mba} \p_\t \psi_i^{\mba} + \f{\l}{4} \hat{\delta}^t_{\mba \mbb\mbc\mbd}  \psi_i^{\mba} \psi_i^{\mbb} \psi_i^{\mbc} \psi_i^{\mbd} \right) \\
& + \int {\rm d}\t  \f{\l \a}{2} \hat{\delta}^t_{\mba \mbb\mbc\mbd}  \left(\psi_1^{\mba} \psi_1^{\mbb} \psi_2^{\mbc} \psi_2^{\mbd} 
+ \psi_1^{\mba} \psi_2^{\mbb} \psi_1^{\mbc} \psi_2^{\mbd} + \psi_1^{\mba} \psi_2^{\mbb} \psi_2^{\mbc} \psi_1^{\mbd}\right) \,,
\end{split}
\ee
where $\hat{\delta}^t_{\mba \mbb\mbc\mbd}$ is the same as in \eqref{eq:deltas}.
The first line represents two copies of the Majorana $O(N)^3$ model introduced in \cite{Klebanov:2016xxf}, and the second line is a coupling between the two, preserving the symmetries and the melonic limit.
Besides the $O(N)^3$ symmetry, the action is invariant under $D_4$, the dihedral group of order 8. In particular, this contains a $\mathbb{Z}_2\times \mathbb{Z}_2$ symmetry that independently flips the sign of the two fields, thus forbidding a mixed two-point function, $G_{12}(\t)=\la \psi_1^{\mba}(\t) \psi_2^{\mba}(0) \ra =0$.
However, Kim et al.\  have shown that for $\a<0$ the bilinear operator $\cO_4^{0} = \psi_1^{\mba} \psi_2^{\mba} -\psi_2^{\mba} \psi_1^{\mba}$ has scaling dimension $h=\f12 + \im f(\a)$, as we will now review, and that the SD equations actually admit in this case a solution with non-vanishing $G_{12}$ having a lower free energy than the symmetric solution. We rephrase here their results in a way that makes explicit the relation with the formalism detailed in Section~\ref{sec:proof}.

One interesting feature of \eqref{eq:S_Kim} is that it allows us to illustrate a case in which the index structure that permeated through our proof of instability plays an important role. In fact, while the instability in the melonic examples we are discussing does not break the $O(N)^3$ invariance, and hence we can concentrate on just the trace part of the fluctuations as in \eqref{eq:diag-ansatz}, we cannot do the same for the flavor indices: in the two-flavor model uncovering the instability requires considering off-diagonal elements $G_{12}(\t_1,\t_2)$.

In order to view the instability of the symmetric solution for $\a<0$ as a particular case of our Proposition~\ref{prop1}, we should first of all address the fact that in this model the fundamental fields are Grassmannian, while in Section~\ref{sec:proof} we worked with bosonic fields.
Nevertheless, it is not difficult to see that the core of the argument still applies: there are of course signs and antisymmetrizations to take care of, in the 2PI effective action and in the basis of bilocal functions, but otherwise the formulas are similar to the bosonic case.
The 2PI effective action for the single-flavor model of \cite{Klebanov:2016xxf} has been written in \cite{Benedetti:2018goh}.
The two-flavor case is slightly more complicated, but the 2PI effective action can be trivially obtained from the action for bilocal variables written in \cite{Kim:2019upg} for the SYK-like version of the model, where the coupling is a random tensor. Such model is defined in terms of $2N$ Majorana fermions $\chi_i^{a}$, with $a=1,\ldots,N$ and $i=1,2$, with the action being given by
\be \label{eq:S_SYK2}
S_{{\rm SYK}^2}[\chi] =  \int {\rm d}\t  \left( \sum_{i=1,2}\left( \f12 \chi_i^{a} \p_\t \chi_i^{a} + \f{1}{4!} J_{a b c d} \, \chi_i^{a} \chi_i^{b} \chi_i^{c} \chi_i^{d} \right)
+\f{6\a}{4!} J_{a b c d} \, \chi_1^{a} \chi_1^{b} \chi_2^{c} \chi_2^{d} \right) \,,
\ee
where $J_{a b c d}$ is a random anti-symmetric tensor with Gaussian distribution, and variance $6 J^2/N^3$.
The path integral of such disordered models can be rewritten in terms of bilocal variables $\tG_{ij}(\t_1,\t_2)=\chi_i^{a}(\t_1) \chi_j^{a}(\t_2)$ which carry only flavor indices (see \cite{Jevicki:2016bwu,Kitaev:2017awl} for the single flavor case). In the new variables, the whole dependence on $N$ factors in front of the action $\tilde{S}_{{\rm SYK}^2}[\tG]$ , hence the large-$N$ limit then reduces to a saddle-point approximation.
In this case, the construction of the 2PI effective action becomes trivial: changing variables to the bilocal ones and taking the large-$N$ limit, the saddle-point approximation gives the functional ${\bf W}[\cJ]$ in \eqref{eq:W[J]} as a Legendre transform of the bilocal action $\tilde{S}_{{\rm SYK}^2}[\tG]$, and since the Legendre transform is an involution, the 2PI effective action is simply $\mathbf{\G}_{{\rm SYK}^2}[G] = \tilde{S}_{{\rm SYK}^2}[G]$.
The result for \eqref{eq:S_SYK2} can be obtained from the one given in \cite{Klebanov:2016xxf}, by setting the Legendre multiplier on shell, $\Sigma_{ij} = \p_\t \d_{ij} - G^{-1}_{ij}$:
\be \label{eq:2PI-SYK2}
\begin{split}
\f{1}{N} \mathbf{\G}_{{\rm SYK}^2}[G] = 
& - \log {\rm Pf} (G^{-1}_{ij}) - \sum_{i=1,2} \int {\rm d}\t \left( \p_\t G_{ii}(\t,\t') \right)_{\t'=\t} - \frac{J^2}{8} \int {\rm d}\t {\rm d}\t' \bigg(\sum_{a,b} G_{ab}^4(\t,\t') \\
&\qquad +6\alpha \Big(G_{12}^2(\t,\t')+G_{21}^2(\t,\t'))(G_{11}^2(\t,\t')+G_{22}^2(\t,\t') \Big) \\
&\qquad +6\alpha^2\Big(G_{11}^2(\t,\t')G_{22}^2(\t,\t') +G_{12}^2(\t,\t') G_{21}^2(\t,\t') \\
& \qquad\qquad +4G_{11}(\t,\t')G_{22}(\t,\t')G_{12}(\t,\t')G_{21}(\t,\t')\Big)\bigg)\,.
\end{split}
\ee
Of course such effective action can also be directly derived from the diagrammatic rules of the 2PI effective action, as done in \cite{Benedetti:2018goh} for the single-flavor SYK model. The advantage of the diagrammatic method is that it works also for tensor models, for which there is no easy change of variables in the path integral.
One thus obtains for \eqref{eq:S_Kim} a 2PI effective action of the same form as \eqref{eq:2PI-SYK2}, with the replacement $J\to \l$.
The rest of the discussion applies equivalently to the models \eqref{eq:S_Kim} and \eqref{eq:S_SYK2}.

The SD equations, obtained from the first variation of \eqref{eq:2PI-SYK2}, admit a solution with $G_{12}=G_{21}=0$, because of the $\mathbb{Z}_2\times \mathbb{Z}_2$ symmetry. For the diagonal components of the two-point function, we have $G_{11}=G_{22}=G$ because of the flavor permutation symmetry, and the SD equations then reduce to
\be
G(\t_{12})= \frac{1}{2}\textrm{sgn}(\t_{12}) + \f{1+3\alpha^2}{2} \l^2N^3\int {\rm d}\t_{3} {\rm d}\t_{4}  \,\textrm{sgn}(\t_{13}) G(\t_{34})^3 G(\t_{42})\,,
\ee
which in the infrared limit is solved by
\be \label{eq:G_star-2flavor}
G_\star(\t) = \left(\frac{1}{4\pi(1+3\alpha^2)}\right)^{\frac{1}{4}}\frac{\textrm{sgn}(\t)}{|\l \t|^{1/2}}\,.
\ee

Next, as in Section~\ref{sec:proof}, we perturb the solution $G_{ij}(\t)=\d_{ij} G_\star(\t)$ in order to study its stability.
The Hessian of the effective action is proportional to $\mathbb{I}_{-}-K$, where now $\mathbb{I}_{-}$ is the identity on the space of bilocal antisymmetric functions.
Arranging the two-point function fluctuations as a vector $(\d G_{11},\d G_{22},\d G_{12},\d G_{21})$, the matrix structure of the Bethe-Salpeter kernel is 
\be
K = 
\begin{pmatrix}
K_{11,11} & K_{11,22} & 0 & 0 \\ 
    K_{22,11} & K_{22,22} & 0 & 0 \\
     0 & 0 & K_{12,12} & K_{12,21} \\
     0 & 0 & K_{21,12} & K_{21,21}
\end{pmatrix}
= 
\begin{pmatrix}
1+\a^2 & 2\a^2 & 0 & 0 \\ 
    2\a^2 & 1+\a^2 & 0 & 0 \\
     0 & 0 & 2\a & 2\a^2 \\
     0 & 0 & 2\a^2 & 2\a
\end{pmatrix}
\f{K_c}{1+3\a^2} \,,
\ee
with 
\begin{align}
K_{c}(\tau_{1},\tau_{2};\tau_{3},\tau_{4}) = -\frac{3}{4\pi} \frac{\textrm{sgn}(\tau_{13})\textrm{sgn}(\tau_{24})}{|\tau_{13}|^{2\Delta}|\tau_{24}|^{2\Delta}|\tau_{34}|^{2-4\Delta}}, \quad \Delta =\frac{1}{4}\,.
\end{align}
The matrix structure is diagonalized by the following eigenvectors:
\be
E^1 = \begin{pmatrix} 1\\ 1\\ 0\\ 0 \end{pmatrix} \,, \;\;
E^2 = \begin{pmatrix} 1\\ -1\\ 0\\ 0 \end{pmatrix} \, \;\;
E^3 = \begin{pmatrix} 0\\ 0\\ 1\\ 1 \end{pmatrix} \,, \;\;
E^4 = \begin{pmatrix} 0\\ 0\\ 1\\ -1 \end{pmatrix} \,.
\ee
The kernel $K_c$ is diagonalized as usual by the three-point conformal structures. However in this case there are two possible conformal three-point structures, antisymmetric and symmetric, respectively, under the exchange of $\t_1$ with $\t_2$:\footnote{We can also think of \eqref{eq:v_A-v_S} as being parity-odd and parity-even (under $\t_i\to-\t_i$), and thus represent a simple instance of what happens when considering fermions in higher dimensions, see \cite{Iliesiu:2015qra}.}
\be \label{eq:v_A-v_S}
v^{A}_{h}(\tau_{1},\tau_{2},\tau_{0}) = \frac{c_A\,\textrm{sgn}(\tau_{12})}{|\tau_{01}|^{h}|\tau_{02}|^{h}|\tau_{12}|^{2\Delta-h}}\,, \quad v^{S}_{h}(\tau_{1},\tau_{2},\tau_{0}) = \frac{c_S\,\textrm{sgn}(\tau_{01})\textrm{sgn}(\tau_{02})}{|\tau_{01}|^{h}|\tau_{02}|^{h}|\tau_{12}|^{2\Delta-h}}\,.
\ee
Since the perturbations must be antisymmetric under simultaneous exchange of field labels and location, one concludes that the complete eigenperturbations (the one-dimensional fermionic version of \eqref{eq:V-def}) are 
\be \label{eq:Vbasis_Kim}
V_{h,\s}(\tau_{1},\tau_{2},\tau_{0}) =
\begin{cases} v^{A}_{h}(\tau_{1},\tau_{2},\tau_{0}) E^i\,, &\quad \s=1,2,3\,,\\
 v^{S}_{h}(\tau_{1},\tau_{2},\tau_{0}) E^4 \,,&\quad \s=4\,,
 \end{cases}
\ee
with eigenvalues
\be
\Big(k_{1}(h), k_{2}(h), k_{3}(h),k_{4}(h) \Big) = \left(g_{A}(h),\frac{1-\alpha^{2}}{1+3\alpha^{2}}g_{A}(h), \frac{2\alpha(1+\alpha)}{1+3\alpha^{2}}g_{A}(h), \frac{6\alpha(1-\alpha)}{1+3\alpha^{2}}g_{S}(h)\right)\,,
\ee
where
\be
g_{A}(h)= -\frac{3}{2}\frac{\tan (\frac{\pi}{2}(h-\frac{1}{2}))}{h-1/2} \,, \quad \; g_{S}(h) =-\frac{1}{2}\frac{\tan (\frac{\pi}{2}(h+\frac{1}{2}))}{h-1/2}\,.
\ee
It should be stressed that since the model is defined in $d=1$, the $V$-functions \eqref{eq:Vbasis_Kim} form a complete basis if and only if we include also the discrete series representations, i.e.\ for $h\in\cP\cup \{h=2n \bigm| n\in \mathbb{N}\}$. As mentioned below Proposition~\ref{prop1}, since $\cP$ is still part of the representation, the inclusion of the discrete series does not affect the instability argument. However, the equation $k_{1}(h)=1$ has a solution at $h=2$, which is on the discrete series. This is a singularity that requires moving away from the conformal limit \cite{Maldacena:2016hyu}, and which plays an important role in the SYK model, but which we will ignore in the present discussion. It could also be removed altogether by considering a long-range version of \eqref{eq:S_Kim}: it is known that in the single-flavor long-range SYK model \cite{Gross:2017vhb} the singularity at $h=2$ is lifted, and the same will happen here, since $k_{1}(h)$ is independent of $\a$.

For $\a<0$, the equation $k_{4}(h)=1$ admits the solutions $h = \f12 \pm f(\a)$, where $f(\alpha)$ solves the equation 
\be
f \tanh (\pi  f/2)  =  -\frac{3\alpha(1-\alpha)}{1+3\alpha^2} \,.
\ee
Such solutions lead to a situation similar to that of Figure~\ref{fig:ON3-plot}. In particular, $1-k_{4}(h)$ has a double pole at $h=1/2$, compensated by the double-zero in the Plancherel weight, which in $d=1$ is \cite{Maldacena:2016hyu,Kitaev:2017hnr}
\be
\r(h) = \f{2h-1}{\pi \tan(\pi h)} \,,
\ee
with a negative limiting value of their product.

Therefore, the two-flavor tensor model \eqref{eq:S_Kim} has an instability for $\a<0$, and the unstable perturbations being in the flavor-mixing sector, $\d G_{12}$ and $\d G_{21}$, they break the discrete $\mathbb{Z}_2\times \mathbb{Z}_2$ symmetry, leaving intact only its diagonal subgroup $\mathbb{Z}_2$ that acts by simultaneously flipping the sign of both fields, $(\psi_1,\psi_2)\to (-\psi_1,-\psi_2)$.
It should be noticed that Kim et al.\ \cite{Kim:2019upg} have gone further than our stability analysis, actually finding numerically another solution of the SD equations and showing that it has a lower free energy than that of \eqref{eq:G_star-2flavor}. Such numerical solution is however peculiar to $d=1$ models, while here we just wanted to show how their result fits in the general picture.

In a more recent paper, Klebanov et al.\ \cite{Klebanov:2020kck} have introduced a similar two-flavor tensor model (and SYK-like counterpart), but with complex fields, and symmetry $SU(N)^2\times O(N) \times U(1)^2$. The $\mathbb{Z}_2\times \mathbb{Z}_2$ symmetry is thus replaced by the continuous abelian group $U(1)^2$ that acts by independently giving a phase factor to each of the two flavors. For $\a<0$ and $\a>1$, one finds again a scaling dimension $h\in \cP$, corresponding to eigenperturbations of the two-point function that break $U(1)^2$, leaving intact only its diagonal subgroup.
And as in the real model, Klebanov et al.\ have been able to show that the SD equation admits a symmetry-breaking solution with a lower free energy than the symmetric one.

%%%%%%%%%%%%%%%%%%%%%%
\subsection{The biscalar fishnet model}
%%%%%%%%%%%%%%%%%%%%%%

The last example we consider is provided by the biscalar fishnet model introduced in \cite{Gurdogan:2015csr}, and studied further in \cite{Grabner:2017pgm,Kazakov:2018qbr,Gromov:2018hut,Gromov:2017cja,Karananas:2019fox,Basso:2019xay} (see also the review \cite{Kazakov:2018ugh}).

Like \eqref{eq:S_Kim}, the fishnet model is also a two-flavor model, but with complex scalar fields in the adjoint representation of $SU(N)$.
In components we write $\phi_i^{ab}$, where $a,b=1,\ldots,N$ are $SU(N)$ indices, and $i=1,2$ is a flavor index. 
The action for general $d\leq 4$ is written as
\be \label{eq:S_fishnet}
\begin{split}
S[\phi] = &N\,\int \dd x \, \Big( \tr
    [\phi_1^\dagger \,\, (- \p^2)^{d/4}\,\phi_1 + \phi_2^\dagger \,\, (-\p^2)^{d/4}\,\phi_2  + (4 \pi)^\frac{d}{2} \xi^2\phi_1^\dagger \phi_2^\dagger \phi_1\phi_2] \Big)\\
    &  +  \,(4 \pi)^\frac{d}{2} \,\int \dd x \, \Big( \alpha_1^2 \, \sum_{i=1}^2\tr[\phi_i\phi_i]\,\tr[\phi_i^{\dagger}\phi_i^{\dagger}]   
       - \alpha_2^2\,\tr[\phi_1\phi_2]\tr[\phi_2^{\dagger }\phi_1^{\dagger }] \\
    &\qquad\qquad\qquad -\alpha_3^2\,\tr[\phi_1\phi_2^{\dagger }]\tr[\phi_2\phi_1^{\dagger }] \Big) \,.
\end{split}
\ee
Besides the $SU(N)$ invariance, the model has also a $U(1)^2$ invariance that acts by independently giving a phase factor to each of the two flavors, as in the two-flavor tensor model of \cite{Klebanov:2020kck}.
There is instead no permutation symmetry between the two fields, because the hermitian conjugate of the single-trace interaction is missing.
It is this chiral nature of the vertex, together with  the planar limit, that leads to a rigid structure of the diagrams, resembling a fishnet, at least in the bulk of large enough diagrams.

Fishnet diagrams are very different from melonic diagrams in general. Nonetheless, as noticed in \cite{Benedetti:2020yvb}, the fishnet model has a number of similarities with the long-range $O(N)^3$ model.
For example, while melonic diagrams characteristically show up in the two-point function, there are no planar fishnet diagrams for the two-point function;
however, for $d<4$ the long-range nature of the kinetic term makes such difference irrelevant from the renormalization point of view, as there is no wave function renormalization in any case.
Moreover, the four-point fishnet diagrams that renormalize the first double-trace terms in \eqref{eq:S_fishnet} are identical to the ladder diagrams generated by the kernel in Figure~\ref{fig:kernel}.
Therefore, one finds a similar structure of beta functions (quadratic in the double-trace couplings, with coefficients depending on the exactly marginal single-trace coupling), fixed points, and spectra of bilinear operators.

Following \cite{Grabner:2017pgm,Kazakov:2018qbr,Gromov:2018hut}, the interesting part of the spectrum of bilinear operators is found by deriving the OPE of the four-point function
$\la \tr[\phi_i (x_1) \phi_i(x_2)] \tr[\phi_i^\dagger (x_3)\phi_i^\dagger(x_4)] \ra$
from the conformal partial wave expansion. As the structure of diagrams contributing to it is the same as in the $O(N)^3$ tensor model, one is led to solve the same Bethe-Salpeter equation for the kernel that we reviewed above, except for a different normalization of the two-point function. With the conventions of \cite{Kazakov:2018qbr}, the kernel eigenvalue is the same as in \eqref{eq:ON3-k}, with the replacement $3g^2/(4\pi)^d\to\xi^4$, and thus the same plots as in Figure~\ref{fig:ON3-plot} are found for real $\xi$.
We conclude that also for the fishnet model with real $\xi$ the conformal vacuum is unstable.

Two questions might be raised about such conclusion. First, we proved Proposition~\ref{prop1} for real fields, while the fishnet model is based on complex fields, so one might wonder if the proposition still applies. It is easy to see that it does, by rewriting \eqref{eq:S_fishnet} as a four-flavor model with real fields, defined by $\phi_1 = (\chi_1 + \im \chi_2)/\sqrt{2}$,  $\phi_2 = (\chi_3 + \im \chi_4)/\sqrt{2}$, with real fields $\chi_i$. The $U(1)^2$ symmetry becomes an $SO(2)^2$ acting as independent rotations on $(\chi_1,\chi_2)$ and $(\chi_3,\chi_4)$. Of course this is not a convenient representation in practice, as in particular the action becomes longer and the $SU(N)$ invariance becomes hidden (the $O(N)$ subgroup remains instead evident), but it shows that in principle it can be treated as a special case of the proof in Section~\ref{sec:proof}.

The second question one might have is how could the 2PI effective action show an instability, 
given that in the planar limit the self-energy of the fishnet model is identically zero, and hence we should also have $\mathbf{\G}_2[G]=0$.
To answer this question, one has to notice that the vanishing of the self-energy relies on the assumption of unbroken $U(1)^2$ symmetry.
Let us first suppose that in constucting the 2PI effective action we couple the bilocal source in \eqref{eq:W[J]} in such a way to respect both the $SU(N)$ and the $U(1)^2$ invariance, i.e.\ we add to the action the term 
\be
S_{\text{symm.}}[\phi,\cJ] = N\int \dd x \dd y\sum_{i=1,2}\cJ_{\bar{i}i}(x,y) \tr[\phi_i^\dagger(x) \phi_i(y)]\,,
\ee
where we use a barred index in the source when this is associated to the hermitian conjugate field. 
After Legendre transform, the 2PI effective action will depend only on two bilocal variables, $G_{\bar{i}i}(x,y)$ for $i=1,2$. Therefore, the diagrammatic rules for the 2PI effective action will be the same as for the original action \eqref{eq:S_fishnet}, and in the planar limit we will find no vacuum diagrams, i.e.\ we will find $\mathbf{\G}_2[G_{\bar{1}1},G_{\bar{2}2}]=0$.
In this case, the SD equations are trivial, just implying that the full two-point functions coincide with the free two-point functions. 
However, such invariant effective action is of limited use,\footnote{If we were to follow the same route for the 1PI effective action, we would in general find it impossible to introduce any source at all. Constructing the effective potential, and thus uncovering possible tachyonic instabilities, generically requires to introduce in the first place a symmetry breaking source.}
and in particular it is impossible to obtain four-point functions such as $\la \tr[\phi_i (x_1) \phi_i(x_2)] \tr[\phi_i^\dagger (x_3)\phi_i^\dagger(x_4)] \ra$ from variations of $\mathbf{\G}[G_{\bar{1}1},G_{\bar{2}2}]$.

Things change if we introduce also symmetry breaking sources, of the type
\be
S_{\text{break.}}[\phi,\cJ] = N\int \dd x \dd y\sum_{i=1,2} \left(\cJ_{ii}(x,y) \tr[\phi_i(x) \phi_i(y)] +  \cJ_{\bar{i}\,\bar{i}}(x,y) \tr[\phi_i^\dagger(x) \phi_i^\dagger(y)] \right)\,.
\ee
With hindsight, we have chosen source terms preserving $SU(N)$ and not mixing $\phi_1$ with $\phi_2$; the only symmetry being broken is the $U(1)^2$ symmetry, which is reduced to $\mathbb{Z}_2{}^2$.
After Legendre transform, the resulting 2PI effective action depends also on $G_{ii}(x,y)$ and $G_{\bar{i}\,\bar{i}}(x,y)$ for $i=1,2$.
The new diagrammatic rules now allow many planar vacuum diagrams, and thus a non-trivial $\mathbf{\G}_2[G]$. However, due to the structure of the vertices, diagrams necessarily have an even number of ``symmetry breaking" propagators, which is what allows the symmetric solution to still be a solution of the field equations:
\be
\begin{split}
&\f{\d \mathbf{\G}_2}{\d G_{\bar{i}i}}\Big|_{G_{ii}=G_{\bar{i}\,\bar{i}}=0}= \f{\d \mathbf{\G}_2}{\d G_{ii}}\Big|_{G_{ii}=G_{\bar{i}\,\bar{i}}=0} = \f{\d \mathbf{\G}_2}{\d G_{\bar{i}\,\bar{i}}}\Big|_{G_{ii}=G_{\bar{i}\,\bar{i}}=0} = 0 \,, \\
& \Rightarrow G^\star_{\bar{i}i}(x,y) = C(x,y)\,, \;\;\; G^\star_{ii}=G^\star_{\bar{i}\,\bar{i}}=0\,.
\end{split}
\ee
On the other hand, the second variations with respect to $G_{ii}$ and $G_{\bar{i}\,\bar{i}}$ 
\be \nn
K_{i\,i\,\bar{i}\,\bar{i}} (x_1,x_2,x_3,x_4) =  -2 \int \dd y_1 \dd y_2 G^\star_{i\bar{i}}(x_1,y_1)G^\star_{i\bar{i}}(x_2,y_2) \f{\d^2 \mathbf{\G}_2[G] }{\d G_{ii}({y_1,y_2}) \d G_{\bar{i}\,\bar{i}}(x_3,x_4)}\Big|_{{G=G^\star}} \,
\ee
is non-vanishing even when evaluated on shell, and it yields precisely the Bethe-Salpeter kernel whose geometric series gives the four-point function $\la \tr[\phi_i (x_1) \phi_i(x_2)] \tr[\phi_i^\dagger (x_3)\phi_i^\dagger(x_4)] \ra$. 
In the large-$N$ limit the only 2PI planar diagrams contributing to $\mathbf{\G}_2[G]$, and having exactly one $G_{ii}$ and one $G_{\bar{i}\,\bar{i}}$ propagator, are a figure-eight with one $\alpha_1^2 \, \tr[\phi_i\phi_i]\,\tr[\phi_i^{\dagger}\phi_i^{\dagger}] $ vertex, and a melon with two single-trace vertices and two more propagators of the type $G_{\bar{i}i}$.
Therefore, the kernel has the same structure as in Figure~\ref{fig:kernel}.
As we anticipated, it is this kernel that admits a solution of the equation $k(h,J)=1$ on the principal series, for real $\xi$ \cite{Grabner:2017pgm,Kazakov:2018qbr,Gromov:2018hut}, and thus
 the model \eqref{eq:S_fishnet} has an instability associated to the perturbations $\d G_{ii}$ and $\d G_{\bar{i}\,\bar{i}}$.

%

%%%%%%%%%%%%%%%%%%%%%%%%%%%%%
\section{Conclusions and outlook}
\label{sec:Outlook}
%%%%%%%%%%%%%%%%%%%%%%%%%%%%%

In this paper we have proved a $d$-dimensional CFT counterpart of the Breitenlohner-Freedman instability in AdS${}_{d+1}$.
The main result is stated in Proposition~\ref{prop1}. Some examples have been discussed in Section~\ref{sec:examples} in order to show some concrete realizations of the setting and formalism used in Section~\ref{sec:proof}. 

As a useful outcome, we now have a proof of instability for tensor and fishnet models in those ranges of parameters leading to a complex scaling dimension belonging to the principal series.
It should be stressed that the instability can be avoided in some ranges of parameters, for example at small coupling in the long-range Amit-Roginsky model, or at imaginary coupling in the long-range $O(N)^3$ tensor model with quartic interaction.\footnote{Another possible cure is to consider supersymmetric versions of these models \cite{Murugan:2017eto,Popov:2019nja,Lettera:2020uay}.} Similarly the dangerous complex scaling dimensions are avoided in the fishnet model for purely imaginary $\xi^2$, although the theory might still be a complex CFT in this case, since for example correlation functions such as $\la \tr[\phi_1{}^L (x_1) ] \tr[\phi_1^\dagger{}^L (x_1) ] \ra$ with odd $L$ receive contributions from ``globe'' fishnet diagrams \cite{Gurdogan:2015csr} with both even and odd number of vertices.
One should also notice that most (but not all \cite{Giombi:2018qgp}) explicit examples with a complex scaling dimension in the principal series share a common feature: an interaction that is unbounded from below. Therefore, in general one expects such models to be unstable at finite-$N$ and the question of existence of a range of parameters in which they are stable to be only relevant in the large-$N$ limit. Unless at the fixed point other (bounded) interactions, such as the pillow and double-trace terms in the $O(N)^3$ tensor model, dominate over the unbounded one, thus leading to a bounded potential even at finite $N$.

Many possible future directions can be envisaged.
First of all, it would be nice to extend the proof in  Section~\ref{sec:proof} to the case of theories whose fundamental fields are fermions (in dimension $d>1$), for which one could use for example some of the machinery developed in \cite{Iliesiu:2015qra,Albayrak:2020rxh}, and possibly even to higher spins. 
Gauge theories should not be difficult to include, as long as the complex dimensions appear in the OPE of the scalar fields.

It would also be interesting to relate the proof to the modern bootstrap perspective \cite{Poland:2018epd}.
Our approach is in fact rather similar in spirit to the so-called ``old bootstrap'' \cite{Polyakov:1970xd,Migdal:1972tk,Parisi:1972zm,Mack:1972kq,Mack:1973mq}, which is also closer to how examples of such situation emerge in tensor and fishnet models, and it shows that such approach can lead to fake CFT solutions, in the sense that sometimes one finds a conformal solution which however is not the stable solution. How does one make sure to avoid such solutions in the modern bootstrap? Of course one can by hand exclude theories with scaling dimensions in the principal series, but not all instabilities of conformal field theories come in this form,\footnote{For example, the Bardeen-Moshe-Bander phenomenon \cite{Bardeen:1983rv} is not associated to complex scaling dimensions.} and it would thus be desirable to have a stability criterion for practical use.

\

One original motivation of the present work was a conjecture advanced in \cite{Kim:2019upg}, that can be stated as following:

\begin{conjecture}\label{conj1} 
Under the same assumptions as in Proposition~\ref{prop1}, in the true vacuum of the theory, the operator $\cO_{h_\star}$ acquires a non-trivial vacuum expectation value: $\Braket{\cO_{h_\star}}\neq 0$.
\end{conjecture}

In other words, the conjecture claims the existence of a stable solution of the SD equations, with spontaneous breaking of conformal invariance.
The instability of the conformal solution was implicitly assumed in \cite{Kim:2019upg}, based on the AdS/CFT picture. With the present work we have completed the picture by presenting a proof that does not rely on the duality. We have instead failed so far to use the same methods in order to prove remaining part of Conjecture~\ref{conj1}.
Ideally one would try to find a stable solution $G_1(x_1,x_2) \neq G_\star(x_1,x_2) $ and show that it leads to $\Braket{ \cO_h}\neq 0$.
However, identifying a stable solution in full generality is out of reach, as it requires going beyond the linear perturbations, hence some new idea would probably be needed.
Proving that a stable vacuum should exist at all seems also non-trivial.
Even in the specific models reviewed  in Section~\ref{sec:examples}, showing that for  $d>1$, in the regime in which complex scaling dimensions appear, there exists a different (and energy-favourable) solution of the SD equations has so far proved to be an elusive task.
Therefore, it cannot even be excluded that in such case those models might have no stable vacuum at all, and that the stable solution found in \cite{Kim:2019upg,Klebanov:2020kck} is peculiar to $d=1$, or to the fermionic nature of the fields.
Looking at the 2PI effective action \eqref{eq:ON3-Gamma_2PI} of the $O(N)^3$ model, it might be tempting to infer that the term proportional to $G^4$ has the good sign only for $\l^2<0$, and that if $\l^2>0$ the effective action is unbounded from below (even at large-$N$) and hence no truly stable vacuum exists. However, the bilocal nature of the effective action makes such a statement not obvious; moreover, in the Amit-Roginsky model we have a stable vacuum at small (real or imaginary) coupling, despite the cubic power of $G$ in the  effective action \eqref{eq:AR-Gamma_2PI}.
It would be desirable to better understand the global stability properties of such bilocal effective actions, and possibly find the true vacuum (if it exists) of some $d>1$ model.

\

Within the framework of Section~\ref{sec:proof}, we can also make a small comment on the distinction between real and complex CFTs.
In  \cite{Gorbenko:2018ncu}, the former have been defined as CFTs having a spectrum of scaling dimensions that are either real or come in complex conjugate pairs (in the second case the CFT is however non-unitary), while the latter can have complex scaling dimensions without their complex conjugate partner.
Examples of real non-unitary CFTs are provided by the Wilson-Fisher fixed point in non-integer dimension \cite{Hogervorst:2015akt} or by the symmetric traceless $O(N)$ model \cite{Jepsen:2020czw}, while an example of complex CFT, with dimensions not on the principal series, is offered by the $Q$-state Potts model with $Q>4$ \cite{Gorbenko:2018dtm}.
In this respect, we notice that a simple corollary of Proposition~\ref{prop1} is the following:

\begin{proposition}\label{prop2} 
Under the assumptions of Hypothesis~\ref{hyp1}, and the first half of Hypothesis~\ref{hyp2}, plus \eqref{eq:Deltas-conditions},
 if the OPE spectrum of two fundamental fields contains a complex scaling dimension, then either it contains also its complex conjugate, or the CFT is unstable.
\end{proposition}

The proof is straightforward. The reality of the kernel implies that the analytic continuations of its eigenvalues satisfy $\overline{k_{\s}(h,J)}=k_{\s}(\bar{h},J)$, hence the set of solutions $\{h_n\}$ of $k_{\s}(h,J)=1$ can only fall into one of the following three cases:
\begin{enumerate}
\item ${\rm Im}(h_n)=0$, $\forall n$, in which case we have a real CFT.
\item There exist also solutions with ${\rm Im}(h_n)\neq 0$, but they all have ${\rm Re}(h_n)> d/2$. From the reality condition it follows that given any such solution, then its complex conjugate $\bar{h}_n$ is also a solution, and since ${\rm Re}(h_n)> d/2$, both poles contribute to the OPE when pushing to the right the contour of the conformal partial wave representation of the four-point function. Therefore, such operators come in conjugate pairs and, in the terminology of \cite{Gorbenko:2018ncu}, the theory is a real non-unitary CFT.
\item There exist also solutions with ${\rm Im}(h_n)\neq 0$ and  ${\rm Re}(h_n)= d/2$, in which case only one solution in the conjugate pair is physical while the other is its shadow. However, Proposition~\ref{prop1} applies and hence the complex CFT is unstable.
\end{enumerate}

Proposition~\ref{prop2} might look rather trivial, as it is essentially saying that a theory with real four-point function has the $\phi\times \phi$ OPE spectrum of a real CFT.
However, it is perhaps useful to state it in such a precise form, and it might help explain why complex CFTs (i.e.\ with complex scaling dimensions not having a complex conjugate partner, and not belonging to the principal series) are not so frequently encountered. 
Moreover, there is at least one non-trivial aspect to be found in the hypotheses. In order to see that, let us consider some possible ways in which Proposition~\ref{prop2} can be violated.
First, the above argument can fail if the kernel eigenvalue does not satisfy the reality condition, which can happen if the Bethe-Salpeter kernel is diagonalizable but not real. In this case, $k_{\s}(h_n,J)=1$ does not imply $k_{\s}(\bar{h}_n,J)=1$.
It can also fail in a less trivial fashion, if the scaling dimensions of the fundamental fields do not satisfy \eqref{eq:Deltas-conditions}. In this case, there could be a physical solution on the left of the principal series, $h_0<d/2$, and then complex solutions might appear after a real solution $h_1>d/2$ merges with the shadow of $h_0$, i.e.\ $\htilde_0=d-h_0>d/2$.\footnote{An example of this kind of merging is found in the long-range sextic $O(N)^3$ tensor model of \cite{Benedetti:2019rja} at imaginary ``wheel" coupling. However, this could be a fake example because $h_1$ crosses marginality (going from $h_1>d$ to $h_1<d$) before reaching the merging, hence the fixed point is probably destabilized by the corresponding dangerously irrelevant operator, in a similar way to what is thought to happen in the long-range Ising model as the power of the Laplacian gets close enough to one \cite{Sak:1973,Behan:2017emf}.} In this case, if the kernel eigenvalue satisfies the reality condition, we would have two pairs of conjugate solutions (as $h_0$ necessarily also merges with $\htilde_1$), but only one in each pair would correspond to a physical operator.
It would be interesting to understand whether this situation can be excluded, and hence if the condition \eqref{eq:Deltas-conditions} can be removed from the hypotheses of Proposition~\ref{prop2}.

A genericity argument also suggests that at finite $N$ it is hard or impossible to find operators with scaling dimension whose real part is exactly $d/2$. In the tensor models for example one finds corrections to the real part at subleading orders \cite{Benedetti:2020sye}, and we expect it to be a general fact.
It would be nice to prove such expectation, or to find a counterexample in which the real part of the scaling dimension is protected from finite-$N$ corrections.

\

Lastly, given the recent construction of an AdS/CFT map for the free or critical $O(N)$ model in \cite{deMelloKoch:2018ivk,Aharony:2020omh}, using very similar techniques to those we used here, it would be interesting to understand the relation between our construction and the proof of the Breitenlohner-Freedman bound in AdS${}_{d+1}$ \cite{Breitenlohner:1982bm,Breitenlohner:1982jf}.

%%%%%%%%%%%%%%%%%%%%%%%%
\section*{Acknowledgements}
\noindent 
The author is grateful to Nicolas Delporte, Razvan Gurau, Igor Klebanov, Slava Rychkov, Kenta Suzuki, and Balt van Rees for useful comments.

%\newpage

%---------------------------------------------
%----- Bibliography ----------------------

\addcontentsline{toc}{section}{References}

\providecommand{\href}[2]{#2}\begingroup\raggedright\endgroup

%---------------------------------------------

%------------------------------------------------------------------------------
\end{document}